\documentclass[a4paper,11pt]{article}
\pdfoutput=1

\usepackage{jheppub1}
\usepackage[T1]{fontenc}
\usepackage{amssymb}
\usepackage{graphicx}
\usepackage{amsmath}
\usepackage{hyperref}
\usepackage{tensor}
\usepackage{epstopdf}
\usepackage{extarrows}
\usepackage{verbatim}
\usepackage{subfigure}
\usepackage{mathrsfs}

\newcommand{\td}{\text{d}}

\begin{document}
	
	\title{On holographic time-like entanglement entropy}
	
\author{Ze Li, Zi-Qing Xiao and Run-Qiu Yang}
	\emailAdd{lize@tju.edu.cn}
    \emailAdd{kingshaw@tju.edu.cn}
	\emailAdd{aqiu@tju.edu.cn}
	\affiliation{Center for Joint Quantum Studies and Department of Physics, School of Science, Tianjin University, Yaguan Road 135, Jinnan District, 300350 Tianjin, P.~R.~China}

	\abstract{In order to study the pseudo entropy of time-like subregions holographically, the previous smooth space-like extremal surface was recently generalized to mix space-like and time-like segments and the area becomes complex value. This paper finds that, if one tries to use  such kind of piecewise smooth extremal surfaces to compute time-like entanglement entropy holographically, the complex area is not unique in general. We then generalize the original holographic proposal of space-like entanglement entropy to pick up a unique area from all allowed ``space-like+time-like'' piecewise smooth extremal surfaces for a time-like subregion. We will give some concrete examples to show the correctness of our proposal.
		
	}
	
	%
	%
	%
	\maketitle
	%
	%
	
\section{Introduction}
The AdS/CFT correspondence says that there is an equivalent relationship between $(d+1)$-dimensional gravity theory in asymptotically anti-de Sitter (AdS) spacetime and $d$-dimensional conformal field theory (CFT) on the boundary of the AdS$_{d+1}$ spacetime. Since it was first proposed in \cite{Maldacena:1997re}, there have been more and more evidences to support it in various different models. In these models people have concluded a few of universal dual relationships between the quantities in gravity theory and boundary field theory. One of the most important applications is holographic entanglement entropy.

For the case that the whole system is described by a pure quantum state, the entanglement entropy of a subsystem is usually computed by the von Neumann entropy formula $S_{vN}=-\rm Tr \rho \rm ln \rho$, where $\rho$ is density matrix of the subsystem. In some situations, the von Neumann entropy is not easy to be calculated directly. However, we can calculate it by the AdS/CFT correspondence. The Ryu-Takayanagi (RT) prescription tells us that the entanglement entropy of boundary CFT can be calculated by codimension-2 extremal surface $\gamma$ in the bulk as~\cite{Ryu:2006bv}
\begin{equation}\label{101}
		S_{A}=\mathop{\rm{ext}}\limits_{\partial \gamma= \partial A}\frac{\rm{Area}(\gamma)}{4G_{N}^{(d+1)}},
\end{equation}
the $\gamma$ extends into bulk and is homologous to subsystem $A$ of dual CFT. The $G_N^{(d+1)}$ is the Newton constant at $(d+1)$-dimensional spacetime. This formula connects the entanglement entropy of dual boundary subsystem and the extremal co-dimensional 2 surface in the bulk. Denote $\gamma_A$ to be a surface which attains the extremal. There are several characteristics of the extremal surface $\gamma_{A}$: (i) The subsystem $A$ is required to be a space-like surface so that it can be encoded quantum state of an ``equal-time'' slice, which leads  that the extremal surface $\gamma_A$ is also space-like. (ii) The extremal surfaces can always be embedded into a co-dimensional 1 Riemannian manifold, where the position of $\gamma_A$ will be determined by 2nd order nonlinear elliptic partial equation with the Dirichlet boundary condition. This implies that the extremal $\gamma_{A}$ are smooth surfaces. (iii) The $\rm{Area}(\gamma_{A})$ is positive real number. In some cases that there are multiple extremal surfaces $\gamma_A$ for a given boundary subregion $A$. The Eq.~\eqref{101} then should be generalized into following form~\cite{Hubeny:2007xt,Rangamani:2016dms}
\begin{equation}\label{generalrt1}
  S_{A}=\text{min} \left\{\text{ext}\lim_{\partial \gamma= \partial A}\frac{\rm{Area}(\gamma)}{4G_{N}^{(d+1)}}\right\}\,,
\end{equation}
i.e. we should take the extremal surface which has the minimum area. This formula gives us unique value of entanglement entropy in terms holography correspondence. Furthermore, if we take the contribution of bulk matters into account, it will require the surface $\gamma$ to minimize the following generalized entropy~\cite{Bekenstein2,PhysRevD.7.2333}
\begin{equation}\label{qes1}
	S_{\rm{EE}}=\text{min} \left\{\text{ext} \left[\lim_{\partial \gamma= \partial A}\frac{\text{Area}(\gamma)}{4G_{N}^{(d+1)}}+S_{\rm bulk}(\Sigma_{\gamma})\right]\right\},
\end{equation}
where $\Sigma_{\gamma}$ is the region surrounded by $\gamma$ and boundary subsystem $A$. and $S_{\rm bulk}(\Sigma_{\gamma})$ is the entropy of matter fields in the region $\Sigma_{\gamma}$. The $\gamma$ that attains Eq.~\eqref{qes1} is called the quantum extremal surface~\cite{Faulkner:2013ana,Engelhardt:2014gca}.

In a static asymptotically AdS spacetime, the extremal surface sits on the canonical time slice and can be Wick rotated into a
Euclidean AdS=CFT setup in a straightforward way. However, such Wick rotation is ambiguous for a time-dependent asymptotically AdS spacetime. The time dependent Euclidean asymptotically AdS space does not correspond to the Wick rotation of a time-dependent asymptotically AdS spacetime. This leads the authors of Ref.~\cite{PhysRevD.103.026005} to consider physical interpretation of extremal surface in time-dependent Euclidean asymptotically AdS space. They found that the area of an extremal surface in a (generically time-dependent) Euclidean asymptotically AdS space calculates the pseudo-entropy~\cite{Guo:2022jzs}.   Pseudo entropy is a generalization of von-Neumann entanglement entropy, which is given by~\cite{Nakata:2020luh,Murciano:2021dga}
\begin{equation}
	S_{PE}=-\rm Tr \tau \rm ln \tau.
\end{equation}
Here $\tau$ is called the transition matrix, which is defined by two states $|\psi\rangle$ and $|\phi \rangle$
\begin{equation}
	\tau=\frac{\left|\psi\right\rangle\left\langle\phi\right|}{\left\langle\phi \mid \psi\right\rangle}.
\end{equation}
Pseudo entropy can be a complex value since the transition matrix $\tau$ is non-Hermitian. It returns to entanglement entropy when we take $|\psi\rangle = |\phi \rangle$. In Euclidean holography, this corresponds to the bulk geometry has time translation symmetry. Properties of pseudo entropy have been studied in~\cite{Mollabashi:2021xsd,Mollabashi:2020yie,Guo:2022jzs}. Note that the holographic value of pseudo entropy in time-dependent Euclidean asymptotically AdS space is still real-valued even though pseudo entropy is complex-valued for general transition matrices.

In previous discussion on holographic entanglement entropy, people focused on the case that the subregion of boundary is space-like. Recently, entanglement entropy of time-like interval has been considered in condensed matter system and holography~\cite{Liu:2022ugc,Doi:2022iyj}. Naively speaking, there will be some troubles in calculating time-like entanglement entropy by formula~\eqref{101}. There is no complete time-like or space-like smooth extremal surface that homologous to the time-like subregion. To connect the boundary of a time-like subregion by extremal surfaces, one should mix both time-like extremal surfaces and space-like extremal surfaces. This fact has a nice holographic interpretation in Ref.~\cite{Doi:2022iyj}:  authors think that the extremal surface is composed of space-like and time-like parts,  which correspond to the real part and the imaginary part of the time-like entanglement entropy respectively.  As a result, the time-like entanglement entropy is found to be a complex value. The entanglement entropy of a time-like interval with width of $T_{0}$ in CFT$_{2}$ is given by~\cite{Doi:2022iyj}
\begin{equation}
	\label{102}
	S=\frac{c_{\mathrm{AdS}}}{3} \ln \left(\frac{T_0}{\epsilon}\right)+\frac{i \pi c_{\mathrm{AdS}}}{6}.
\end{equation}
This result can be obtained by performing analytical continuation based on the space-like entanglement entropy. Using the trick of analytical continuation, Ref.~\cite{Doi:2022iyj} also obtains complex-valued entanglement entropies for time-like intervals in BTZ black hole and higher dimensional AdS spacetime. It was argued that such time-like entanglement
entropy and holographic entanglement entropy in dS/CFT~\cite{Maldacena:2002vr,Narayan:2015vda,Sato:2015tta,Narayan:2022afv} are also  related with each other via an analytical continuation~\cite{Doi:2022iyj,Narayan:2022afv}.

In Ref.~\cite{Doi:2022iyj}, the complex-valued pseudo entropy is also holographically interpreted in terms of a group of particular piecewise smooth space-like extremal surfaces and time-like extremal surfaces. If the holographic dual of time-like entanglement entropy is a well-defined quantity, it should be given without referring to the analytical continuation of space-like entanglement entropy. From the viewpoint of practice, the analytical continuation method is not so convincing, and it is not good at dealing with the more general case. Since to perform the analytical continuation, we should obtain the explicit expression of entanglement entropy for arbitrary space-like intervals. When the dual boundary does not correspond vacuum CFT, there is no such explicit analytical expression in general. Using the holographic ``space-like+time-like'' extremal surface that mentioned in Ref.~\cite{Doi:2022iyj} looks like a good way to calculate the time-like entanglement entropy. However, we will use detailed computations to show that the area of ``time-like+space-like'' extremal surface is not unique even for models considered in Ref.~\cite{Doi:2022iyj}. In fact, there are always infinitely many ``time-like+space-like'' extremal surfaces for a given time-like subregion, which give us infinitely many different complex-valued areas. Authors of Ref.~\cite{Doi:2022iyj} found only particular such combinations which coincide with the results of analytical continuations. However, if we do not know the results of analytical continuations, how do we choose the right one from infinitely many different complex-valued areas? To answer this question is our main task in this paper. Our idea is to generalize the definition of extremal surface, which we will call ``complex-valued weak extremal surface'', so as to pick up the unique area from all possible ``space-like+time-like'' extremal surfaces and can give the right time-like entanglement entropy. Of course, when we return to the space-like subsystems, complex-valued weak extremal surface returns to the traditional space-like extremal surface and gives the correct entanglement entropy by~\eqref{101}.

The organization of this paper is as follow. In the Sec.~\ref{section 2} we will first use the example of AdS$_3$-CFT$_2$ duality to show that the ``time-like+space-like mixed'' extremal surface is not unique in general for a given timelike subregion and we can obtain infinitely many different complex-valued areas. Then we will give our proposal about how to generalize the holographic entanglement entropy for timelike subregions. In the Sec.~\ref{examp1} and Sec.~\ref{moreexamples} we examine our proposal in various different situations and show that our proposal just gives us the results as same as analytical continuations. A short summary and discussion can be found in Sec.~\ref{dis}.

\section{Complex-valued weak extremal surface}\label{section 2}

\subsection{Time-like entanglement entropy in AdS$_{3}$/CFT$_{2}$}
In Ref.~\cite{Doi:2022iyj}, the authors have studied the entanglement entropy of a time-like interval in the case of  AdS$_{3}$/CFT$_{2}$. Let us first simply review the relevant discussions in Ref.~\cite{Doi:2022iyj}. Consider the Poincar\'{e} patch of a AdS$_3$ spacetime, of which the metric reads
\begin{equation}\label{ads3}
  \td s^2=R_{\text{AdS}}^2\frac{\td z^2-\td t^2+\td x^2}{z^2}\,.
\end{equation}
Here $R_{\text{AdS}}$ stands for the AdS radius. The time-like interval $A$ on the boundary CFT$_{2}$ has finite width $T_{0}$ in time direction and fixed spatial coordinate, i.e. $A=\{t\in[-T_0/2,T_0/2],x=x_0\}$. The entanglement entropy of $A$, which is called time-like entanglement entropy, is given by a complex value
\begin{equation}
	\label{201}
	S_{A}=\frac{c_{\mathrm{AdS}}}{3} \ln \left(\frac{T_0}{\epsilon}\right)+\frac{i \pi c_{\mathrm{AdS}}}{6}.
\end{equation}
Here $c_{\mathrm{AdS}}=\frac{3R_{\mathrm{AdS}}}{2 G_{\mathrm{N}}}$ is the central charge of CFT$_{2}$, $\epsilon$ is UV cut-off.  This result can be obtained by analytical continuation referring to the space-like entanglement entropy. The  space-like entanglement entropy is given by $S=\frac{c}{3} \ln \left(L/\epsilon\right)$~\cite{Calabrese:2004eu}, where $L$ is the length of the space-like interval. We can make a replacement $L \rightarrow iT_{0}$ and then obtain the  time-like entanglement entropy~\eqref{201}. In addition, the time-like entanglement entropy can be interpreted as pseudo entropy when we consider it in the analytical continuation way~\cite{Doi:2022iyj}. Analytic continuation leads to non-Hermitian and complex value entropy, which are the properties of pseudo entropy.
\begin{figure}
\centering
\includegraphics[width=0.3\textwidth]{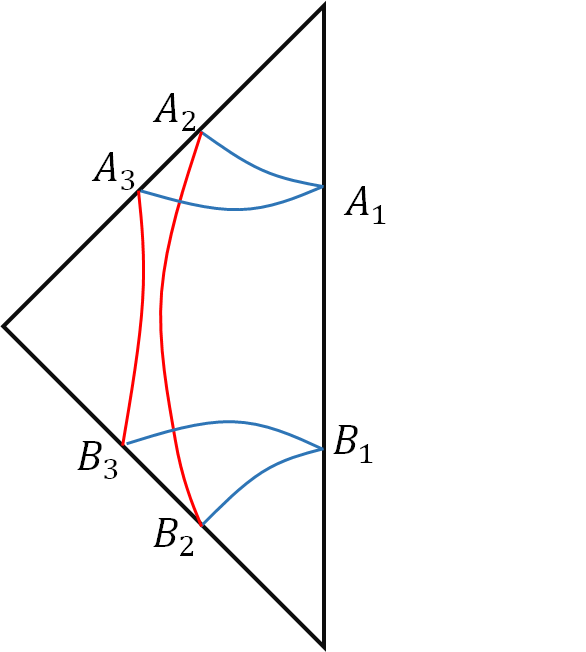}
\includegraphics[width=0.3\textwidth]{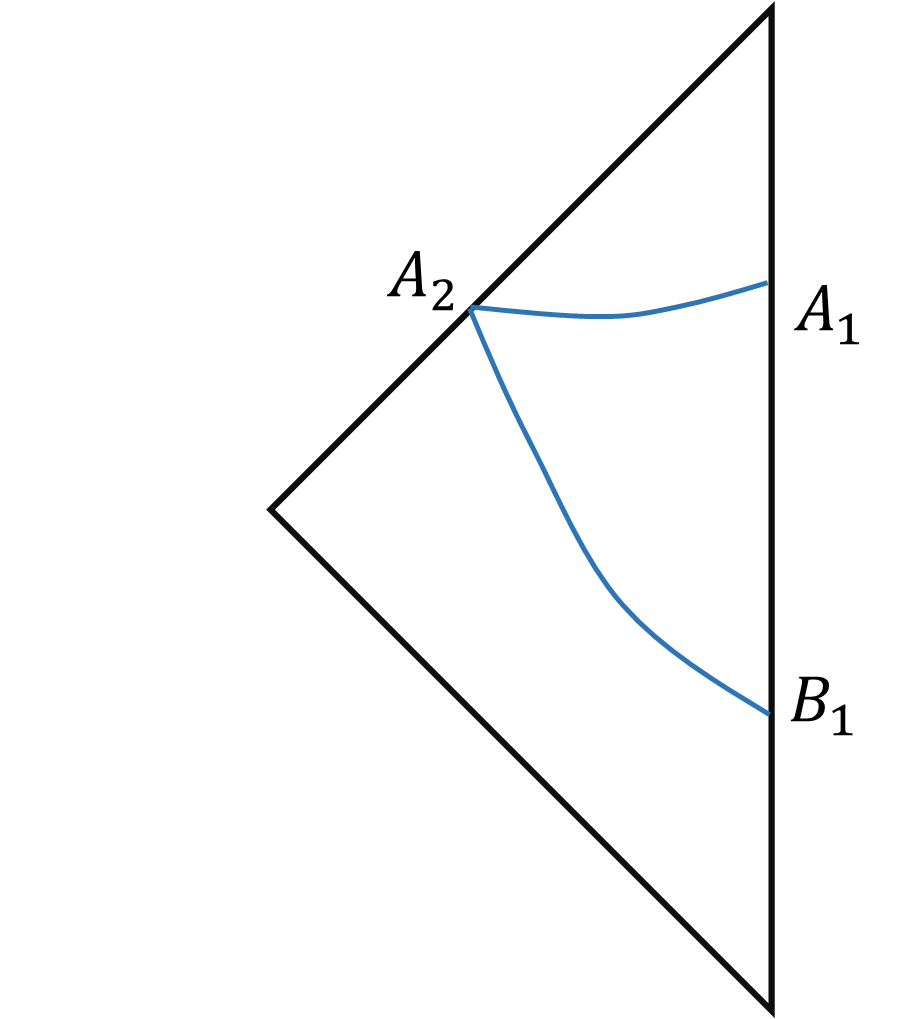}
   \caption{Penrose diagrams on Poincar\'{e} patch of AdS$_3$ spacetime and geodesics connecting $\partial A$. Left panel: The blue solid curves $A_1A_2$ and $B_1B_2$ stand for the space-like geodesics used in Ref.~\cite{Doi:2022iyj} and the red solid curve $A_2B_2$ stands for the time-like geodesic. The  blue solid curves $A_1A_3$ and $B_1B_3$ and the red solid curve $A_3B_3$ stand for other possible geodesics. Right panel: Using two space-like geodesics to connect the endpoints of time-like interval $A$. }\label{figads3}
\end{figure}

To understand the result~\eqref{201}, Ref.~\cite{Doi:2022iyj} then gives a holographic deduction. Due to the lack of space-like geodesics connecting $\partial A$, Ref.~\cite{Doi:2022iyj} argues that we should uses two space-like geodesics to connect the endpoints of $\partial A$ and null infinities, then use a time-like geodesic to connect other two endpoints of two space-like geodesics. One can see the left panel of schematic diagram~\ref{figads3}.  The time-like interval $A$ is denoted by segment $A_1B_1$. The blue solid curves $A_1A_2$ and $B_2B_2$ stand for the space-like geodesics used in Ref.~\cite{Doi:2022iyj}, which connecting the endpoints of $\partial A$ and future/past null infinities respectively. The red curve $A_2B_2$ stands for the time-like geodesic used in Ref.~\cite{Doi:2022iyj}. Holographic duality of time-like entanglement entropy in AdS$_{3}$ is given by the length of these three geodesics (1-dimensional extremal surface). The geodesics $A_1A_2$ and $B_1B_2$ are given by~\cite{Doi:2022iyj}
\begin{equation}
	\label{203}
	t=\pm\sqrt{z^2+T_0^2 / 4},~~z\in(0,\infty)
\end{equation}
in the Poincar\'{e} coordinate, which is a hyperbola with two branches. It is space-like geodesic, whose total length is given by
\begin{equation}
	\label{204}
	S=\frac{R_{\mathrm{AdS}}}{4 G_{\mathrm{N}}} \cdot 2 T_0 \int_\epsilon^{\infty} \frac{d z}{2z \sqrt{z^2+T_0^2 / 4}}=\frac{c_{\mathrm{AdS}}}{3} \ln \frac{T_0}{\epsilon}.
\end{equation}
This result is exactly the real part of~\eqref{201}. Two branches of hyperbola go to $(t=+\infty, z=+\infty)$ and $(t=-\infty, z=+\infty)$ respectively. The geodesic connecting the two points $A_2$ and $B_2$ is a time-like geodesic $A_2B_2$ (the red solid curve), which formally can be given by following equation
\begin{equation}\label{timeb1b2}
  z=\sqrt{(t-t_0)^2+z_0^2},~~t\in(-\infty,\infty)\,.
\end{equation}
Here $t_0$ and $z_0$ are two constants, which can be determined by the positions of endpoints $A_2$ and $B_2$. Though the position of such time-like geodesic depends on the value of $t_0$ and $z_0$, its length is constant and just gives the imaginary part of~\eqref{201}.

The above is the conclusion in Ref.~\cite{Doi:2022iyj}. The authors give the time-like entanglement entropy through analytic continuation, and give the explanation of its holographic dual. Since we have obtained the right complex-valued pseudo entropy for the time-like interval $A$ by analytical continuation, the above holographic interpretation sounds good. However, let us forget the result of analytical continuation and consider the above holographic computation.  One can find that there are some ambiguities in the above holographic computation.

The equation of the geodesic is $z\ddot{z}+\dot{z}^2-1=0 $, where dot represents derivative with respect to $t$. The general solution of spacelike geodesic is given locally by
\begin{equation}
	\label{205}
	t=\pm \sqrt{z^2+C_{1}^2}+C_{2}.
\end{equation}
The general solution has two branches, and each branch has two undetermined constants $C_1$ and $C_2$. If we choose two branches with the same constants $C_1=T_0/2$ and $C_2=0$, we will get~\eqref{203} and~\eqref{204}. However, there is no mathematical or physical requirement that we have to do such choice. To match the endpoints of $\partial A$, the two branches of geodesics in general could be
\begin{equation}
	\label{206}
		t=\sqrt{z^2+C_{+}^2}-C_{+}+T_{0}/2,~~t=-\sqrt{z^2+C_{-}^2}+C_{-}-T_{0}/2\,.
\end{equation}
The $C_{+}$ and $C_{-}$ are two free parameters. Without losing generality, we here take $C_\pm$ to be two nonnegative constant. Choosing different $C_+$ and $C_-$ will give us different space-like geodesics (see the blue and red solid curves $A_1A_3B_3B_1$ of Fig.~\ref{figads3}). \footnote{We will see later that there is only one independent freedom of parameters $\{C_+, C_-\}$. } We can finally get the sum of two geodesic branch lengths
\begin{equation}
	\label{207}
	S=\frac{c_{\mathrm{AdS}}}{6} \ln \frac{4C_{+}C_{-}}{\epsilon^2}.
\end{equation}
If we agree that the space-like geodesic will go to future infinity $(t=+\infty, z=+\infty)$ and past infinity $(t=-\infty, z=+\infty)$, the time-like geodesic will the same as that in Ref.~\cite{Doi:2022iyj}. The length of time-like geodesic $\pi c_{\mathrm{AdS}}/6$ gives the imaginary part of time-like entanglement entropy. The total length of time-like and space-like geodesics is
\begin{equation}
	\label{208}
	S=\frac{c_{\mathrm{AdS}}}{6} \ln \frac{4C_{+}C_{-}}{\epsilon^2}+\frac{i \pi c_{\mathrm{AdS}}}{6},
\end{equation}
which is a complex number. Without referring to the analytical continuation, how do we determine the values of $C_+$ and $C_-$ so that they can also give us the result~\eqref{201}?

Noting the fact that every a pair of $C_\pm$ stands for a possible ``space-like+time-like'' geodesic pair to connect the $\partial A$,  naively, one may say we should choose the $C_+$ and $C_-$ such that the total length of space-like geodesics Eq.~\eqref{207} becomes minimal, just as what we do in computing the entanglement of space-like interval holographically: if there are multiple extremal surfaces, we then choose the one of minimal area. Note that we can choose the value $C_+\rightarrow0^+$. This in fact gives an almost null geodesic. We will see later that, in this case, the value of $C_-$ will approach into a finite positive value. Then one can see that the value of Eq.~\eqref{207} can be infinitely negative and so has no lower bound. Thus, we cannot apply mechanically the experience of space-like case. In addition, once one admits piecewise smooth extremal surfaces to replace the smooth extremal surface, it seems that more natural choice would be to use two space-like geodesics to connect the endpoints of time-like interval $A$, see the right panel of Fig.~\ref{figads3}. Without referring to the result of analytical continuation of space-like interval, why do we ignore this configuration?

\subsection{Extremal surface with complex valued area}
In order to give our answer to above questions, we first need to generalize the definition of extremal surfaces. As mentioned above, there may be extremal surface with time-like part and space-like part mixing. We should better redefine the area of the surface so that it can take complex values. The traditional area of codimensional-2 surface is
\begin{equation}
	\label{209}
	\mathrm{Area}=\int \sqrt{|h|} \td x^{d-1},
\end{equation}
where $h$ is determinant of inducted metric on the codimensional-2 surface. The absolute value under the square root ensures that the area is a non negative real number. Now we defined the complex valued area as
\begin{equation}
	\label{210}
	\mathscr{A} =\int \sqrt{h} \td x^{d-1}.
\end{equation}
Obviously, for a time-like surface, $\mathscr{A}$ will be a pure imaginary number, while for a space-like surface, $\mathscr{A}$ will be a real number. There will always be $\rm{Im}(\mathscr{A}) \geqslant 0$ and $\rm{Re}(\mathscr{A}) \geqslant 0$.
If a surface is smooth, it is either a time-like surface or a space-like surface. However, there is no such extremal surface for timelike subregions in the boundary of an asymptotically AdS spacetime. If we admit piecewise smooth surfaces which mix spacelike and timelike segments, we need to determine the extremal value of the area of in these piecewise smooth surfaces. Since the area may be complex number, the ``minimal'' or ``maximal'' is ambiguous and we should defined a relationship between complex numbers to compare with each other.\\
\textbf{Definition: } \textit{For two different complex numbers $z_{1}$ and $z_{2}$, we call $z_1$ to be ``larger than'' $z_2$ and denote $z_{1}\succ z_{2} $, if}
\begin{equation}
	\label{213}
	\left\{
	\begin{aligned}
		&\text{Im}(z_{1})> \text{Im}(z_{2})\,,\\
		&\text{or Im}(z_{1}) = \text{Im}(z_{2}), \text{Re}(z_{1})> \text{Re}(z_{2})\,.
	\end{aligned}
	\right.
\end{equation}
\textit{The ``smaller'' relationship $z_1\prec z_2$ is also defined in a similar way. In addition, for a complex numbers set U, we call $z_0\in U$ to be maximum (minimum) if no element of U is larger (smaller) than $z_0$. }

Let us denote $X_{\mathcal{B}}$ to the set of all co-dimensional-2 piecewise smooth Lipschitz continuous surfaces which share the same boundary $\mathcal{B}$.
Strictly speaking, to make our following discussion complete mathematically, we should first specify a topology $\mathcal{T}$ for the set $X_{\mathcal{B}}$ properly so that $X_{\mathcal{B}}$ becomes a topological space. The detailed choice of $\mathcal{T}$ is not relevant on our following discussions and we will not go further on this subtle issue. Piecewise smooth surface $\Gamma$ can be written as $\Gamma=\bigcup_{i}^{N}\Gamma_{i}$, where $\{ \Gamma_{i} \}$ are smooth surfaces, $N$ is a finite positive integer. The joint ``points'' of two piece of smooth surfaces can be marked as $E_{ij}=\Gamma_{i}\cap \Gamma_{j}$ and we call it a ``jointed edge''. The traditional RT extremal surface can be generalized to piecewise smooth surfaces, and we call it \itshape complex-valued weak extremal surface\upshape (CWES) in this paper. Its definition is as follow:\\
\textbf{Definition: } \itshape We call $\Gamma\in X_{\mathcal{B}}$ to be a CWES if following conditions are satisfied:
\begin{enumerate}
\item[(1)] Its every smooth segment $\Gamma_i$ is an extremal surface;
\item[(2)] If $\Gamma_i$ is not the unique extremal surface enclosed by $\partial\Gamma_i$, then $\Gamma_i$ is the one of minimal area in these extremal surfaces;
\item[(3)] Under the requirements (1) and (2), the $\mathscr{A}(\Gamma)$ will be functional of jointed edges. Then if we deformed the jointed edges infinitesimally, the $\mathscr{A}(\Gamma)$ will be local maximum or minimum in the sense of ``$\succ$'' and ``$\prec$''.
\end{enumerate}
\upshape
If the $\Gamma$ has only one smooth segment, then one can realize that requirements (1) and (2) just give us the usual RT surface. When there are two or more smooth segments, the requirement (3) then plays role. Denote $E_{ij}$ to be one jointed edge of $\Gamma$. In the case that functional derivative $\delta\mathscr{A}(\Gamma)/\delta E_{ij}$ is well defined, we can obtain some more explicit expressions for requirement (3). If the jointed edge $E_{ij}$ is inside the manifold $\mathcal{M}$,  the property (3) then requires
\begin{equation}\label{deltagammab}
		\frac{\delta \mathscr{A}(\Gamma)}{\delta E_{ij}}=0\,.
	\end{equation}
If the $E_{ij}$ is in a boundary $T$ of $\mathcal{M}$, we then should use
\begin{equation}\label{deltagammab2}
		\left.\frac{\delta \mathscr{A}(\Gamma)}{\delta E_{ij}}\right|_T=0\,.
	\end{equation}
to replace Eq.~\eqref{deltagammab}. Here the script $T$ means that the variation $\delta E_{ij}$ should keep the ``joint points'' $E_{ij}$ in the boundary $T$. Note that Eqs.~\eqref{deltagammab} and \eqref{deltagammab2} both contain the equations for real and imaginary parts.

\begin{figure}
\centering
\includegraphics[width=0.4\textwidth]{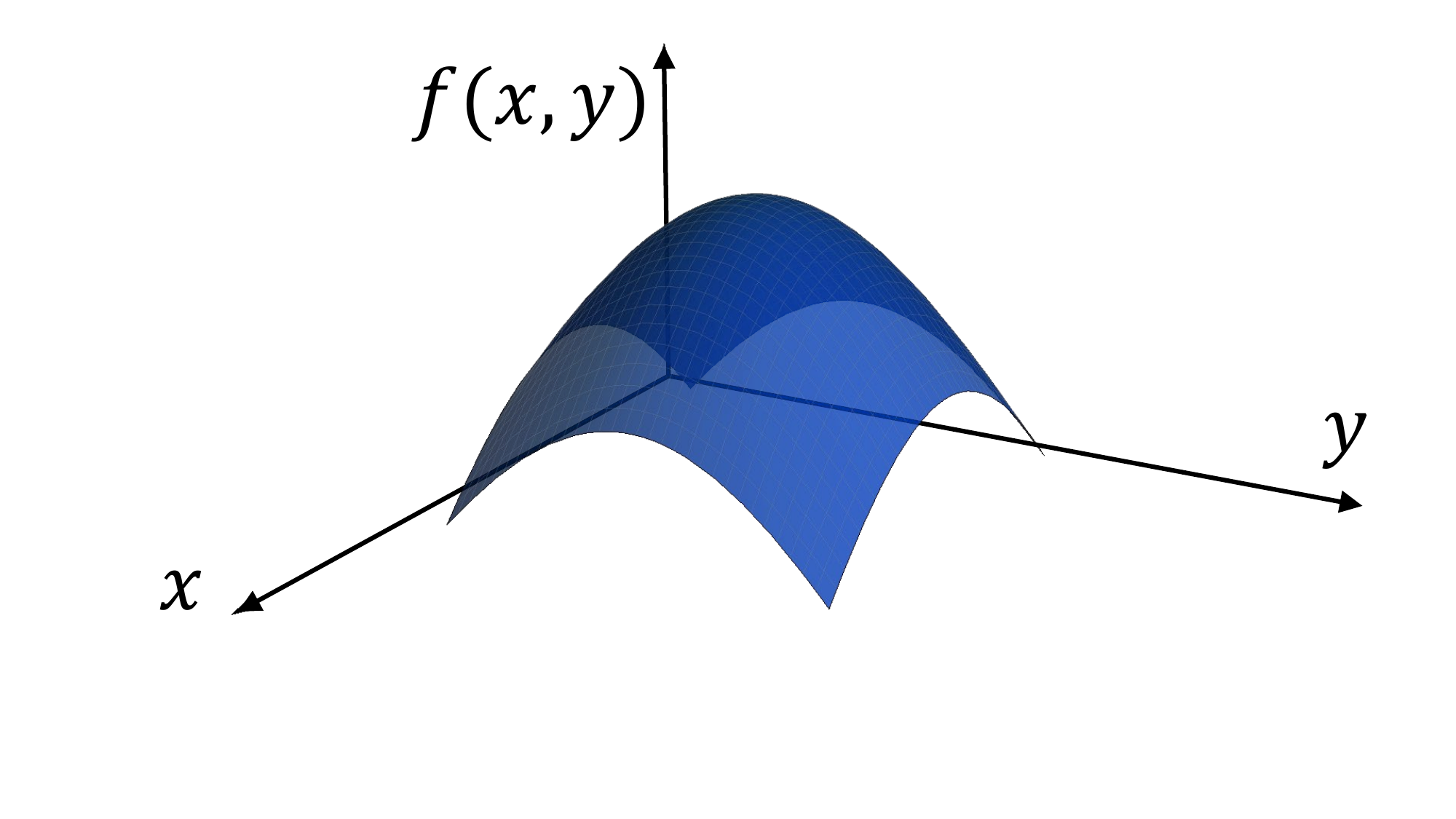}
\includegraphics[width=0.4\textwidth]{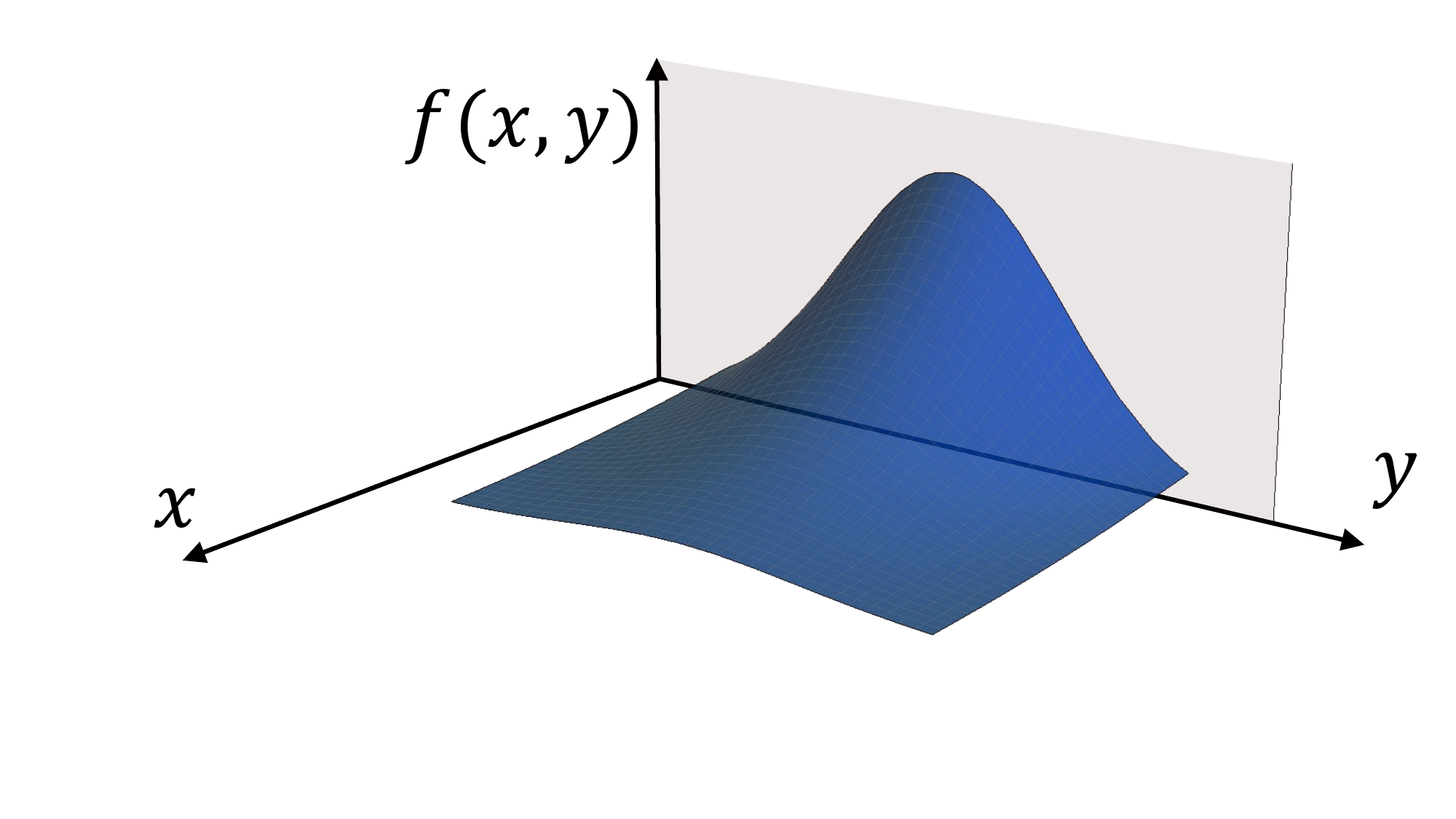}
   \caption{The analog of multiple variables function for the local extremal points. When the maximal point is in the interior of definition domain, we require $\partial_xf=\partial_yf=0$, i.e. $\td f=0$. However, it the maximal point locates at boundary of definition domain, we only needs $\partial_yf=0$ but $\td f\neq0$ in general. }\label{twodfun}
\end{figure}

Let us now explain a little more on why we need to distinguish the above two different situations. Let us consider the analog of multiple-variable function $f(x,y)$. The definition domain of $f(x,y)$ is $y\in\mathbb{R}$ and $x\geq0$. See the schematic diagram~\ref{twodfun}.
For point $E=(x_0,y_0)$ with $x_0>0$, it is clear that $E$ is one maximal value will imply
\begin{equation}\label{detalfb1}
  \left.\frac{\td}{\td s}f(x_0+sv_1,y_0+sv_2)\right|_{s=0}=0, ~~\forall (v_1,v_2)\in\mathbb{R}^2\,.
\end{equation}
This is an analog of Eq.~\eqref{deltagammab} and is equivalent to
\begin{equation}\label{detalfb1a}
  \left.\frac{\partial}{\partial x}f\right|_E=\left.\frac{\partial}{\partial y}f\right|_E=0\,.
\end{equation}
However, if the point $E=(0,y_0)$, i.e. is belong to a point of boundary, the necessary condition to insure $E$ to be local maximum is only
\begin{equation}\label{detalfb1b}
  \left.\frac{\partial}{\partial y}f\right|_E=0\,.
\end{equation}
The equation
\begin{equation}\label{detalfb2}
  \left.\frac{\partial}{\partial x}f\right|_E=0\,
\end{equation}
is NOT necessary. In this case we can also replace Eq.~\eqref{detalfb1} by
\begin{equation}\label{detalfb1c}
  \left.\frac{\td}{\td s}f(x_0,y_0+sv_1)\right|_{s=0}=0, ~~\forall v_1\in\mathbb{R}\,.
\end{equation}
We see that only the derivative along the boundary is necessary to be zero in this case. This is the analog of Eq.~\eqref{deltagammab2}.
If one embeds the manifold $\mathcal{M}$ into a larger space $\bar{\mathcal{M}}$, a CWES $\Gamma$ may not be a CWES any more. This is because $T$ is no longer the boundary and the lager space $\bar{\mathcal{M}}$ may admit some new linear independent variations of $E_{ij}$ which may break the requirement (3). One also notes that Eqs.~\eqref{deltagammab} and \eqref{deltagammab2} are two necessary conditions of requirement (3) due to two reasons: (i) the functional derivative may not exist, and (ii) some ``saddle points'' that satisfy Eqs.~\eqref{deltagammab} and \eqref{deltagammab2} may not correspond to local maximum or minimum.
%

With this definition, it is possible to obtain more than one CWES. In the original RT proposal, one should take the minimal area in multiple extremal surfaces. We here generalize holographic entanglement entropy formula \eqref{generalrt1} into following form
\begin{equation}
	\label{214}
	S_{A}=\mathrm{Min}\left\{\mathop{\rm{Ext}}\limits_{\partial \Gamma= \partial A}\frac{\mathscr{A}(\Gamma)}{4G_{N}^{(d+1)}}\right\}.
\end{equation}
Here ``$\mathrm{Min}$'' stands for the minimum value defined by relationship ``$\succ$'', ``$\mathscr{A}$'' represents the complex valued area and ``Ext'' means that we find the complex-valued weak extremal surfaces. This formula can return to the space-like entanglement entropy due to the definition of ``$\succ$''.\footnote{This is the main reason why we use Eq.~\eqref{213} to define ``$\succ$''. One may use an alternative way to define ``$\succ$'', i.e. we first compare the real part and then compare the imaginary part. Note that even for a spacelike subregion $A$, we may still find some piecewise extremal surfaces which also have boundary $\partial A$ but mix spacelike and timelike segments. If one uses the alternative way to define ``$\succ$'', one cannot  guarantee the spacelike smooth extremal surface always has smallest real part and so cannot guarantee the spacelike subregion always have real entanglement entropy.} Currently, we cannot prove this formula, but we will check it with the following examples.

\section{Example of AdS$_{3}$/CFT$_{2}$}\label{examp1}
Before we go further to discuss our generalization of holographic time-like entanglement entropy, let us first consider again the most basic example discussed in Ref.~\cite{Doi:2022iyj}. Now we do not refer any assistance of analytical continuation and directly consider the question based on our holographic propose. We first consider the case that $\mathcal{M}$ is the static patch covered by Poincar\'{e} coordinates and then consider the case that $\mathcal{M}$ is the covered by global coordinates.

\subsection{Poincar\'{e} patch}
\begin{figure}
\centering
\includegraphics[width=0.9\textwidth]{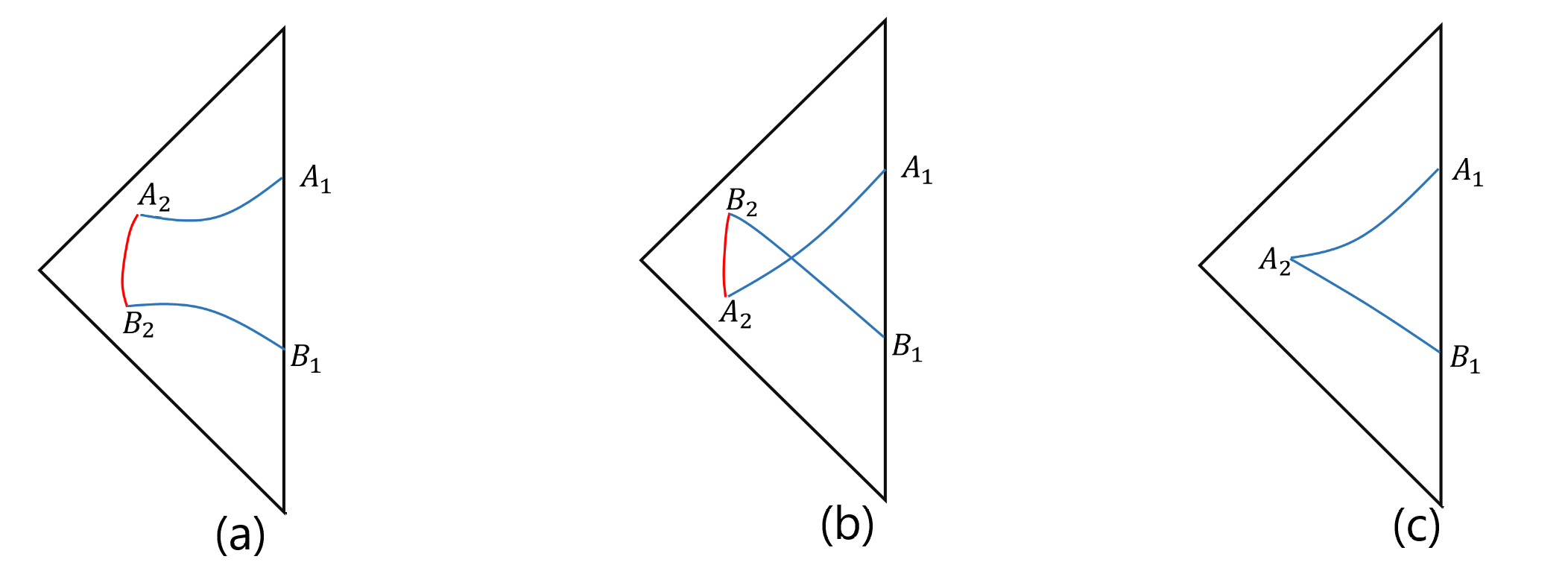}
   \caption{There are three different possible configurations to connect the endpoints of time-like interval $A$ by a few of geodesics. }\label{figads3b}
\end{figure}

\begin{figure}
\centering
\includegraphics[width=0.25\textwidth]{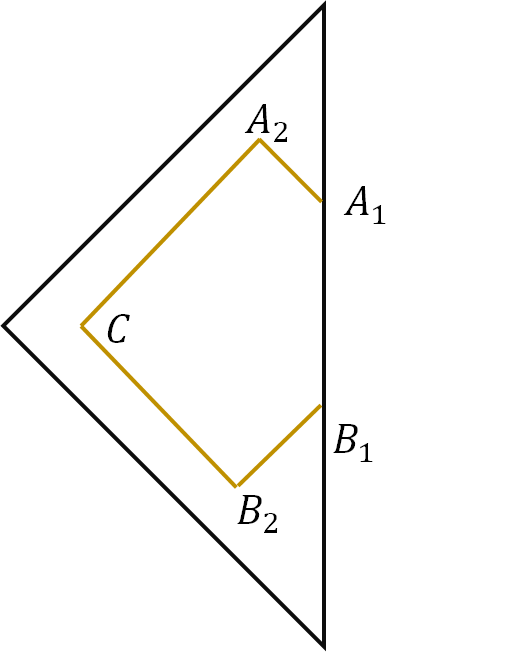}
\includegraphics[width=0.25\textwidth]{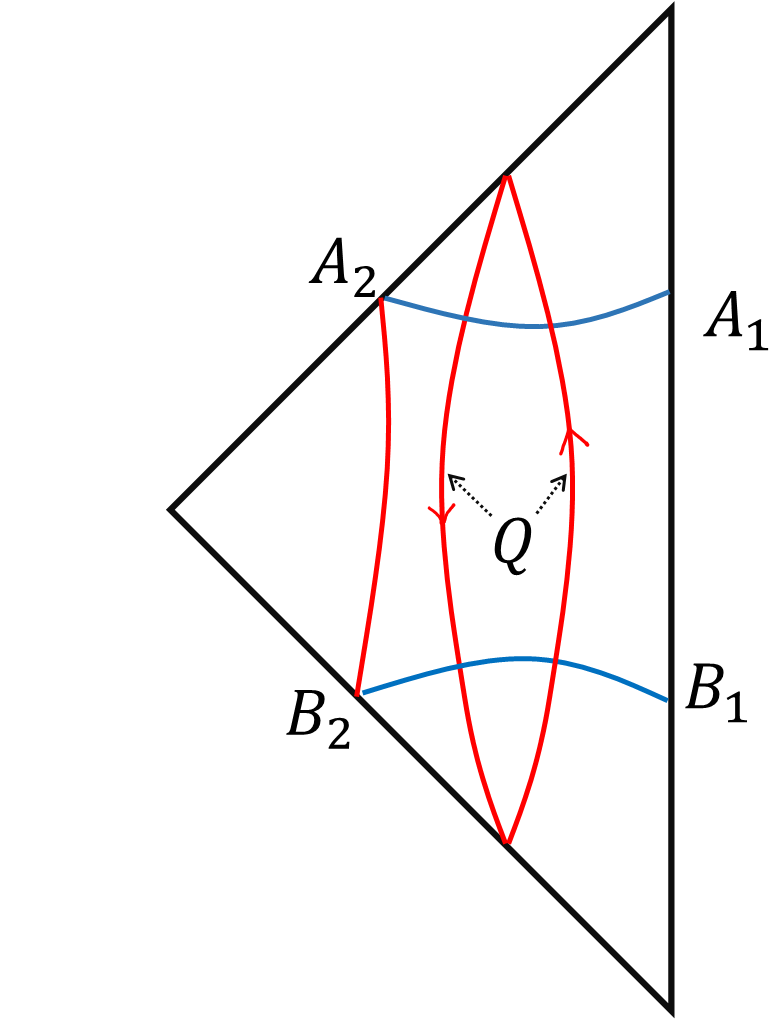}
   \caption{Left: For a timelike interval in a Poincar\'{e} patch  of AdS$_{3}$ spacetime, we can always use four null geodesics to connect its endpoints. Such piecewise null geodesics has infinitely many different choices but they all have zero area. Right: the CWES contains a connected branch $A_1A_2B_2B_1$ and a closed timelike branch $Q$. }\label{fignull}
\end{figure}

For the Poincar\'{e} patch of AdS$_2$, we choose the coordinates~\eqref{ads3}. Since there is no time-like geodesic arrive at AdS boundary, we cannot use a smooth geodesic to connect the two endpoints of time-like interval $A$. The the CWES must be contains a few of piecewise smooth extremal surfaces. Due the translation symmetry along $x$ direction, our proposal then implies that the holographic time-like entanglement entropy must be given by one of the three different relevant configurations shown in the schematic diagram of Fig.~\ref{figads3b} or the configuration shown in the left panel of Fig.~\ref{fignull}. In the configurations (a) (b) and (c) of Fig.~\ref{figads3b}, the blue solid curves stand for space-like geodesics and red solid curve stand for time-like geodesics.  Differing from Ref.~\cite{Doi:2022iyj} that directly assume the joint points of time-like geodesic and space-like geodesic are in null infinities, here we assume that these joint points may be in the general positions of bulk. Now let us analyze them one by one.

Firstly, let us consider the configuration (c) of Fig.~\ref{figads3b}, where we use two different space-like geodesics $A_1A_2$ and $A_2B_1$ to connect the endpoints of time-like interval $A_1B_1$. Based on our rule, the length of this configuration, if it is a CWES, will be smaller than other two since it has no imaginary part. This directly puts our proposal in danger since configuration (c) impossibly gives us the correct result~\eqref{201}. However, we in following can show that configuration (c) is not a CWES and so our proposal just excludes the configuration (c).

Let us parameterize the two space-like geodesics $A_1A_2$ and $B_1A_2$ by the Eq.~\eqref{206}. Then the position of $A_2=(z_2,t_2)$ can be obtain the combination of two equation in  Eq.~\eqref{206}. The total length of these two geodesics is
\begin{equation}\label{confcl1}
  \mathscr{A}=R_{\text{AdS}}\int_\epsilon^{z_2} \frac{C_+\td z}{z \sqrt{z^2+C_+^2}}+R_{\text{AdS}}\int_\epsilon^{z_2} \frac{C_-\td z}{z \sqrt{z^2+C_-^2}}
\end{equation}
Here $z_2$ is the function of $C_\pm$. The condition~\eqref{deltagammab} or \eqref{deltagammab2} then requires that $\partial \mathscr{A}/\partial C_\pm=0$, which leads to
\begin{equation}\label{confcl2}
   \left[\frac{C_+}{z_2 \sqrt{z_2^2+C_+^2}}+\frac{C_-}{z_2 \sqrt{z_2^2+C_-^2}}\right]\frac{\partial z_2}{\partial C_{\pm}}=0\,.
\end{equation}
However, from Eq.~\eqref{206} we see
\begin{equation}\label{confcl3}
  \sqrt{z_2^2+C_+^2}+\sqrt{z_2^2+C_-^2}=C_-+C_+-T_0\,.
\end{equation}
This gives us following result
\begin{equation}\label{confcl4}
  \left[\frac{z_2}{\sqrt{z_2^2+C_+^2}}+\frac{z_2}{\sqrt{z_2^2+C_-^2}}\right]\frac{\partial z_2}{\partial C_{\pm}}=1-\frac{C_{\pm}}{\sqrt{z_2^2+C_{\pm}^2}}>0\,.
\end{equation}
It is clear that Eqs.~\eqref{confcl4} and \eqref{confcl2} cannot be both satisfied. We conclude that total length given by Eq.~\eqref{confcl1} is not extremal and our proposal excludes the configuration (c).

Let us now consider the configurations (b). The spacelike geodesics are described by
\begin{equation}\label{geodb1}
\begin{split}
  &A_1A_2:~~t=-\sqrt{z^2+C_+^2}+C_++T_0/2\\
  &B_1B_2:~~t=\sqrt{z^2+C_-^2}-C_--T_0/2\,.
\end{split}
\end{equation}
Here $C_\pm$ are assumed to be nonnegative. The time-like extremal surface is still described as
\begin{equation}\label{timeb1b2c}
  z=\sqrt{(t-t_0)^2+z_0^2}\,.
\end{equation}
Let us now denote the positions of $A_2$ and $B_2$ are
$$A_2=(z_2,t_2),~~~B_2=(z_3,t_3),~~t_3>t_2\,.$$
From Eq.~\eqref{timeb1b2c} we read the imaginary part of $\mathscr{A}$
\begin{equation}\label{confa2}
  \text{Im}\mathscr{A}=R_{\text{AdS}}\int_{t_2}^{t_3} \frac{z_0\td t}{(t-t_0)^2+z_0^2}=R_{\text{AdS}}\left[\arctan\left(\frac{t_3-t_0}{z_0}\right)-\arctan\left(\frac{t_2-t_0}{z_0}\right)\right]\,.
\end{equation}
Treat the total length $\mathscr{A}$ to be the function of position of $A_2$ and $B_2$, we see that the Eq.~\eqref{deltagammab} requires
\begin{equation}\label{confa30}
  \frac{\partial}{\partial t_2}\text{Im}\mathscr{A}= \frac{\partial}{\partial t_3}\text{Im}\mathscr{A}=0\,.
\end{equation}
This leads to $t_3=-t_2=\infty$. Now we expand Eq.~\eqref{timeb1b2c} and the  second one of Eq.~\eqref{geodb1} near the future null infinity and obtain following two equations
\begin{equation}\label{nulltz1}
  t=z-T_0/2-C_-+\mathcal{O}(1/z),~~z=t-t_0+\mathcal{O}(1/t)\,.
\end{equation}
These two equations should give us the same point $A_2$ when $z\rightarrow\infty$. We then conclude that $t_0=-T_0/2-C_-$. For the similar reason we also have $t_0=T_0/2+C_+$. Then  the two parameters $C_+$ and $C_-$ are not independent and we have
\begin{equation}\label{relct0}
 C_-+C_+=-T_0<0\,.
\end{equation}
This is contradictory to the requirement $C_\pm\geq0$. Thus, Eq.~\eqref{confa30} has no solution and the configuration (b) cannot be a CWES.

We now consider the configuration (a). The two spacelike geodesics in this configuration are given by Eq.~\eqref{206}. Denote the positions of $A_2$ and $B_2$ to be
$$A_2=(z_2,t_2),~~~B_2=(z_3,t_3),~~t_2>t_3\,.$$
We then have
\begin{equation}\label{confa1}
  \text{Re}\mathscr{A}=R_{\text{AdS}}\int_\epsilon^{z_2} \frac{C_+\td z}{z \sqrt{z^2+C_+^2}}+R_{\text{AdS}}\int_\epsilon^{z_3} \frac{C_-\td z}{z \sqrt{z^2+C_-^2}}
\end{equation}
The time-like geodesic $A_2B_2$ is still given by Eq.~\eqref{timeb1b2c}. We have found the joint points of CWES must be at null infinities, so the imaginary of $\mathscr{A}$ reads
\begin{equation}\label{confa2}
  \text{Im}\mathscr{A}=R_{\text{AdS}}\left.\arctan\left(\frac{t-t_0}{z_0}\right)\right|_{-\infty}^{\infty}=\pi R_{\text{AdS}}\,.
\end{equation}
Now we expand Eq.~\eqref{timeb1b2c} and the first one of Eq.~\eqref{206} near the future null infinity and obtain following two equations
\begin{equation}\label{nulltz1}
  t=z+T_0/2-C_++\mathcal{O}(1/z),~~z=t-t_0+\mathcal{O}(1/t)\,.
\end{equation}
These two equations should give us the same point $A_2$ when $z\rightarrow\infty$. We then conclude that $t_0=T_0/2-C_+$. For the similar reason we also have $t_0=-T_0/2+C_-$. Then we have
\begin{equation}\label{relct0}
 C_-+C_+=T_0\,.
\end{equation}
Then we find the real part of $\mathscr{A}$
\begin{equation}\label{confa3}
  \text{Re}\mathscr{A}=R_{\text{AdS}} \ln \frac{4C_{+}C_{-}}{\epsilon^2}=R_{\text{AdS}} \ln \frac{4C_{+}(T_0-C_+)}{\epsilon^2}.
\end{equation}
To be a CWES we have to set $\partial \mathscr{A}/\partial C_+=0$, which shows that $C_+=T_0/2$. The two space-like segments are given by
\begin{equation}\label{ads3gamma1}
		A_1A_2: ~t=\sqrt{z^2+T_0^2/4},~~B_1B_2:~t=-\sqrt{z^2+T_0^2/4},~~z\in(0,\infty)\,.
\end{equation}
and total length reads
\begin{equation}\label{totalAads3}
  \mathscr{A}=2R_{\text{AdS}}\ln\frac{T_0}{\epsilon}+iR_{\text{AdS}}\pi\,.
\end{equation}
This gives us the correct result~\eqref{102}.

Except for the common configurations shown in the Fig.~\ref{figads3b}, there is also an other special configuration shown in the left panel of Fig.~\ref{fignull}: we can use four null geodesics to connect endpoints $A_1$ and $B_1$. Since we admit the piecewise smooth timelike and spacelike extremal surfaces, there is no prior reason to naively exclude such a configuration.
If such kind of configuration is a CWES, then it will have zero area. Based on our proposal, this will be the one of minimal area and so puts our proposal in danger again. However, we can shown that this is not a CWES and so our proposal just excludes this configuration. Let us consider the null geodesic $A_1A_2$, which can be parameterized by
\begin{equation}\label{nulla1a2}
  t=z+T_0/2\,.
\end{equation}
Now let us consider an infinitesimal deformation on the jointed edge $A_2$, which deform the geodesic $A_1A_2$ into a spacelike geodesic and Eq.~\eqref{nulla1a2} becomes
\begin{equation}\label{nulla1a2b}
  t=\sqrt{z^2+\delta C^2}-\delta C+T_0/2\,.
\end{equation}
Here $\delta C$ is an infinitesimal positive number. The length of this geodesic then reads
\begin{equation}\label{nulla1a2c}
  \mathscr{A}(A_1A_2)=R_{\text{AdS}}\ln\frac{\delta C}{\epsilon}\,.
\end{equation}
For the infinitesimal deformation, it is clear that $\delta\mathscr{A}/\delta C\rightarrow\infty$ and so this configuration is not a CWES. We see that we here only used our holographic proposal to obtain the unique and correct complex area but did not refer to any result of analytical continuation.

In above discussion, we only discussed the simple connected configurations. When we take the piecewise smooth extremal surfaces into account, it is also interesting to consider the multiply connected configurations. Suppose that a closed timlike curve $Q$ contains two smooth timelike geodesics and touches the two null boundaries of Poincar\'{e} patch. See the right-panel of Fig.~\ref{fignull}. Though the closed timlike curve $Q$ is not a smooth timelike geodesic, it is a CWES. Since $\partial Q=\emptyset$, we see that $A_1A_2B_2B_1\cup Q$ is still a CWES enclosed by the $\partial A_1B_1$. Thus, the most general CWES will have following form
\begin{equation}\label{generacwes1}
  \text{(simply connected CWES)}\cup Q_1\cup Q_2\cdots\cup Q_n
\end{equation}
Here $\{Q_1, Q_2,\cdots, Q_n\}$ stands for different closed timelike curves, every one of which contains two smooth timelike geodesics and touches the two null boundaries of Poincar\'{e} patch. As the result, we obtain the complex-value lengths
\begin{equation}\label{totalAads3b1}
  \mathscr{A}_n=2R_{\text{AdS}}\ln\frac{T_0}{\epsilon}+iR_{\text{AdS}}(2n+1)\pi,~~~n=0,1,2,\cdots\,.
\end{equation}
This gives us the entropy
\begin{equation}\label{totalAads3e1}
 S_n=\frac{c_{\text{AdS}}}6\left[2\ln\frac{T_0}{\epsilon}+(2n+1)\pi i\right],~~~n=0,1,2,\cdots\,.
\end{equation}
Based on our Eq.~\eqref{214}, we then obtain
\begin{equation}\label{totalAads3f1}
 S=\text{Min}S_n=\frac{c_{\text{AdS}}}6\left(2\ln\frac{T_0}{\epsilon}+\pi i\right)\,.
\end{equation}
Let us consider the analytical continuation of spacelike entanglement entropy. For a general spacelike interval with endpoints $(0,0)$ and $(X,T_0)$, if it is a spacelike interval. i.e. $X^2>T_0^2$, the entropy then is given by length of smooth spacelike geodesic,
\begin{equation}\label{totalAads3f21}
 S=\frac{c_{\text{AdS}}}6\ln\frac{X^2-T_0^2}{\epsilon^2}\,.
\end{equation}
We now analytically continue this expression into the case $X^2<T_0^2$ and so
\begin{equation}\label{totalAads3f2}
 S=\frac{c_{\text{AdS}}}6\ln\frac{X^2-T_0^2}{\epsilon^2}\xlongequal[]{X=0}\frac{c_{\text{AdS}}}6\left[2\ln\frac{T_0}{\epsilon}+(2n+1)\pi i\right]\,.
\end{equation}
Interestingly, the result~\eqref{totalAads3e1} just corresponds to the multiple values \eqref{totalAads3f2} when we analytically continue logarithmic function into complex plane.

\subsection{Global AdS$_3$}
\begin{figure}
\centering
\includegraphics[width=0.8\textwidth]{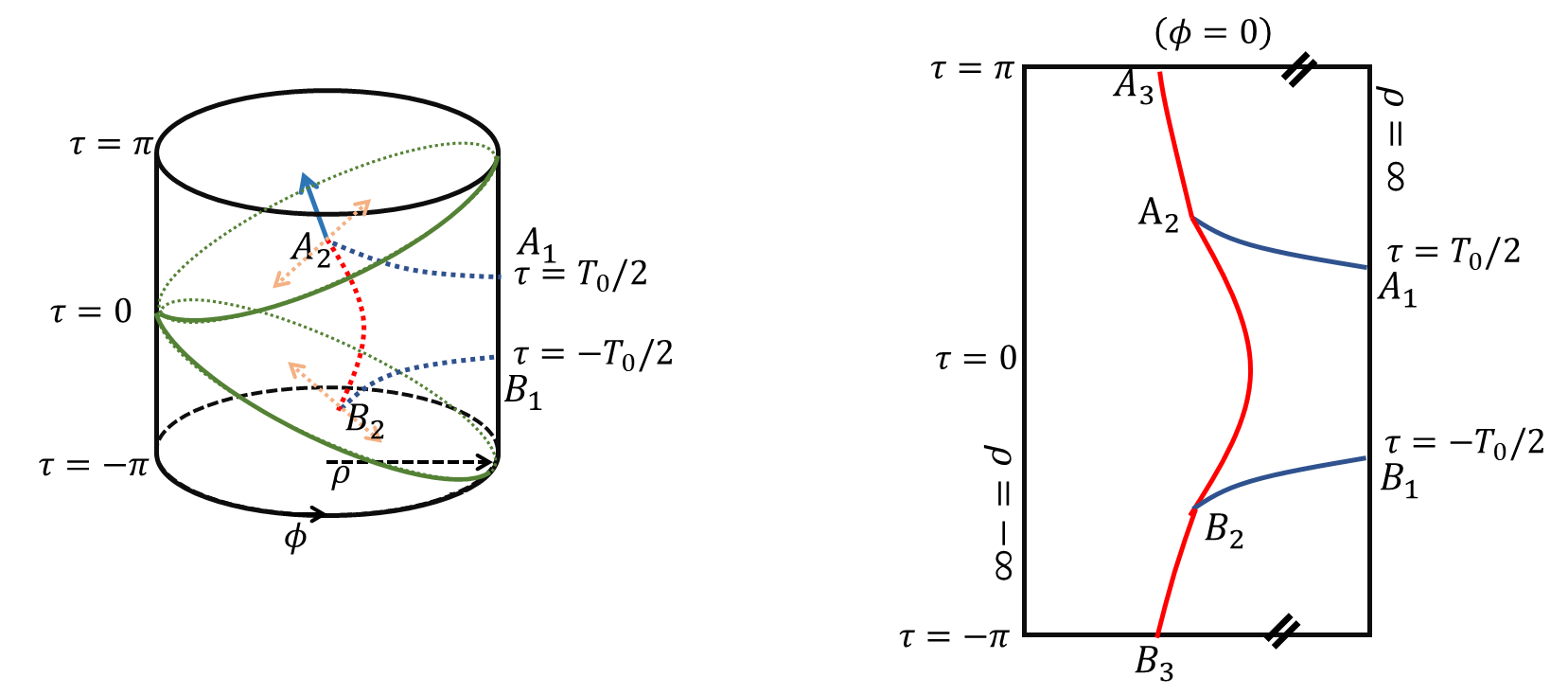}
   \caption{The global coordinate of AdS$_3$ spacetime. The green boundary stands for the null boundary of Poincar\'{e} patch. The blue and red dashed curves stand for the CWES of Poincar\'{e} patch. The right panel is the cut of $\phi=0$. The blue and red solid curves stand for the spacelike and timelike geodesics. Note that the global AdS$_3$ spacetime has topology $S^1\times R^2$, of which the time coordinate is periodic $\tau\sim\tau+2\pi$. Thus, the point $A_3$ and $B_3$ are same point and curve $A_2A_3B_3B_2A_2$ forms a closed circle. }\label{figads3c}
\end{figure}

We now consider the global AdS$_3$. Due the maximally symmetry of AdS spacetime, for the time-like interval we can always choose above coordinate suitably such that the time-like interval is given by
$$A:~~~\phi=0,~\tau\in[-T_0/2, T_0/2]\,,$$
and the metric is
\begin{equation}\label{global1}
  \td s^2=R_{\text{AdS}}^2\left[-(1+\rho^2)\td\tau^2+\frac{\td\rho^2}{1+\rho^2}+\rho^2\td\phi^2\right]
\end{equation}
The Penrose diagram of global AdS$_3$ is shown in Fig.~\ref{figads3c}. Note that the global AdS$_3$ spacetime has topology $S^1\times R^2$, of which the time coordinate is periodic by $\tau\sim\tau+2\pi$. Due to this reason, we can restrict
\begin{equation}\label{restrict0}
  T_0\in(0,\pi]\,.
\end{equation}
In the left panel of Fig.~\ref{figads3c} we show the diagram of global AdS$_3$ spacetime and its Poincar\'{e} patch. The null infinities of Poincar\'{e} patch is denoted by green curves. In general, when we enlarge the original manifold by analytical continuation, the original CWES may not be a CWES of the new manifold if some points of original CWES locate at the boundary of original manifold. In the left panel of Fig.~\ref{figads3c}, the Poincar\'{e} patch is bounded by two null boundaries (denoted by green solid and dashed curves). The spacelike segments and timelike segment of CWES in Poincar\'{e} patch are denoted by blue dashed curves and red dashed curves. The joint points of spacelike and timelike segments locate at null boundaries. Restricted in the Poincar\'{e} patch, the infinitesimal variation of joint points can only move along the null boundary (indicated by orange dashed arrows of Fig.~\ref{figads3c}), which gives us $\delta\mathscr{A}=0$. However, when we embed the Poincar\'{e} patch into globally AdS$_3$ spacetime,  we can variant the joint point along $\tau$-direction (denoted by blue arrows). This variation of joint points can give us $\delta\mathscr{A}\neq0$. Thus, the CWES of Poincar\'{e} patch may be no longer the CWES of global AdS$_3$ spacetime and we cannot simply use the result of Poincar\'{e} patch.

Let us now find the CWES for global AdS$_3$ spacetime. Assume that the spacelike segments of CWES is given by the blue geodesics $A_1A_2$ and $B_1B_2$ shown the right panel of Fig.~\ref{figads3c}. Due to the symmetry, we can assume that $A_2=(\rho_0,\tau_0)$ and $B_2=(\rho_0,-\tau_0)$ with $\tau_0>0$. Since the time direction is closed, there is always two timelike geodesic to connect $A_2$ and $B_2$. The one is the red curve $A_2B_2$ and the other one is the red curve $A_2A_3B_3B_2$. Note that $A_3$ and $B_3$ are same point. If points $A_2$ and $B_2$ meet each other, the CWES becomes piecewise spacelike geodesics and they cannot be extremal since a spacelike curve has extremal length only if it is a smooth spacelike geodesic. When $A_2\neq B_2$, we can separate $A_2$ and $B_2$ farther and farther, then Im$\mathscr{A}$ becomes larger and larger and so $\mathscr{A}$ becomes larger and larger (in the sense of ``$\succ$''). Since the timelike geodesic always have two candidates (the $A_2B_2$ and $A_2A_3B_3B_2$) and we should choose the smaller one (based on the requirement (2) of our definition on CWES), one can realize that the imaginary part of $\mathscr{A}$ becomes extremal only if $\tau_0=\pi/2$, where the lengths of $A_2B_2$ and $A_2A_3B_3B_2$ have same length. To see this more clearly, we parameterize timelike geodesic by $\rho=\rho(t)$. Then the geodesic equation reads
\begin{equation}\label{adsgeoeq1}
  \ddot{\rho}=-\frac{\rho^4+2\rho^2-3\dot{\rho}^2+1}{1+\rho^2}\rho\,,
\end{equation}
of which the solution reads

\begin{equation}\label{soluads31}
  \rho=\frac{c_1\cos t}{\sqrt{1-c_1^2\cos^2t}},~~t\in[-\tau_0,\tau_0],~~0<\tau_0\leq\pi/2\,.
\end{equation}

Here $c_1$ is arbitrary real number which satisfies $0<c_1<1$. When $\tau_0>\pi/2$, we have to choose the branch $A_2A_3B_3B_2$ and obtain

\begin{equation}\label{soluads31}
  \rho=\frac{c_1\cos t}{\sqrt{1-c_1^2\cos^2t}},~~t\in[\tau_0,\pi)\cup(-\pi,-\tau_0],~~\pi>\tau_0>\pi/2\,.
\end{equation}

The length of shorter timelike geodesic in these two situations can be combined into following formula
\begin{equation}\label{lengthtimeads3}
  \text{Im}\mathscr{A}=2\arctan\left(\frac{|\tan\tau_0|}{\sqrt{1-c_1^2}}\right),~~\tau_0\in(0,\pi)\,.
\end{equation}
Treat the $\tau_0$ as the parameter and we see that there is no solution $\partial_{\tau_0}\text{Im}\mathscr{A}=0$ but has a non-differentiable point at $\tau_0=\pi/2$ which attains local maximum. This means that the complex area $\mathscr{A}$ attains local maximum by our definition. Thus, according to our definition, the CWES must appear at the configuration of $\tau_0=\pi/2$. We see the endpoint $A_2$ and $B_2$ the located at $\rho_0=0$ and $\tau_0=\pm\pi/2$. This gives us the length of timelike geodesic in the CWES
\begin{equation}\label{timeads31}
  \text{Im}\mathscr{A}=R_{\text{AdS}}\pi\,,
\end{equation}
which is independent of $c_1$. For the spacelike geodesics, we parameterize them by $t=t(\rho)$ and obtain following equation
\begin{equation}\label{ads3rhot1}
  \tau''=(\rho^2+1)\rho\tau'^3-\frac{3\rho\tau'}{\rho^2+1}\,.
\end{equation}
Here the prime stands for the derivative with respective to $\rho$. This gives us two different types of spacelike geodesics
\begin{equation}\label{ads3space1}
  \text{type I: }\tau(\rho)=\pm\left[\arctan\left(\frac{\rho}{\sqrt{c_2^2\rho^2+1+c_2^2}}\right)+T_0/2-\arctan\frac1{c_2}\right]
\end{equation}
and
\begin{equation}\label{ads3space2}
  \text{type II: }\tau(\rho)=\pm\left[-\arctan\left(\frac{\rho}{\sqrt{c_2^2\rho^2+1+c_2^2}}\right)+T_0/2+\arctan\frac1{c_2}\right]
\end{equation}
with $c_2>0$. The ``$+$'' gives us the geodesic $A_1A_2$ and ``$-$'' gives us the geodesic $B_1B_2$. Only the type II geodesic can reach $\rho=0$ and $\tau=\pi/2$ since we restrict $T_0\leq\pi$. The length of type II spacelike geodesic reads
\begin{equation}\label{lengthad31}
  \text{Re}\mathscr{A}=2R_{\text{AdS}}\ln\left[\frac{2c_2}{c_2\rho_0+\sqrt{c_2^2\rho_0^2+c_2^2+1}}\right]-2R_{\text{AdS}}\ln\epsilon\,.
\end{equation}
%
It passes through the endpoint $\rho_0=0$ and $\tau_0=\pi/2$, so we have
\begin{equation}\label{ads3space1}
  T_0/2+\arctan\frac1{c_2}=\pi/2\Rightarrow c_2=\tan\frac{T_0}{2}\,.
\end{equation}
This determines the unique the desired spacelike part of CWES and we obtain

\begin{equation}\label{lengthad31}
  \text{Re}\mathscr{A}=2R_{\text{AdS}}\ln\frac{2c_2}{\sqrt{c_2^2+1}}-2R_{\text{AdS}}\ln\epsilon=2R_{\text{AdS}}\ln\left(\frac{2\sin T_0/2}{\epsilon}\right)\,.
\end{equation}
Since the length of real part is unique, these timelike and spacelike geodesics give us the maximum length (in the sense of ``$\succ$'') and we then find the ``area'' of the CWES
\begin{equation}\label{totalAads3b}
  \mathscr{A}=2R_{\text{AdS}}\ln\frac{2\sin(T_0/2)}{\epsilon}+iR_{\text{AdS}}\pi\,.
\end{equation}
This is the result obtain by Ref.~\cite{Doi:2022iyj} by using analytical continuation.

\begin{figure}
\centering
\includegraphics[width=0.6\textwidth]{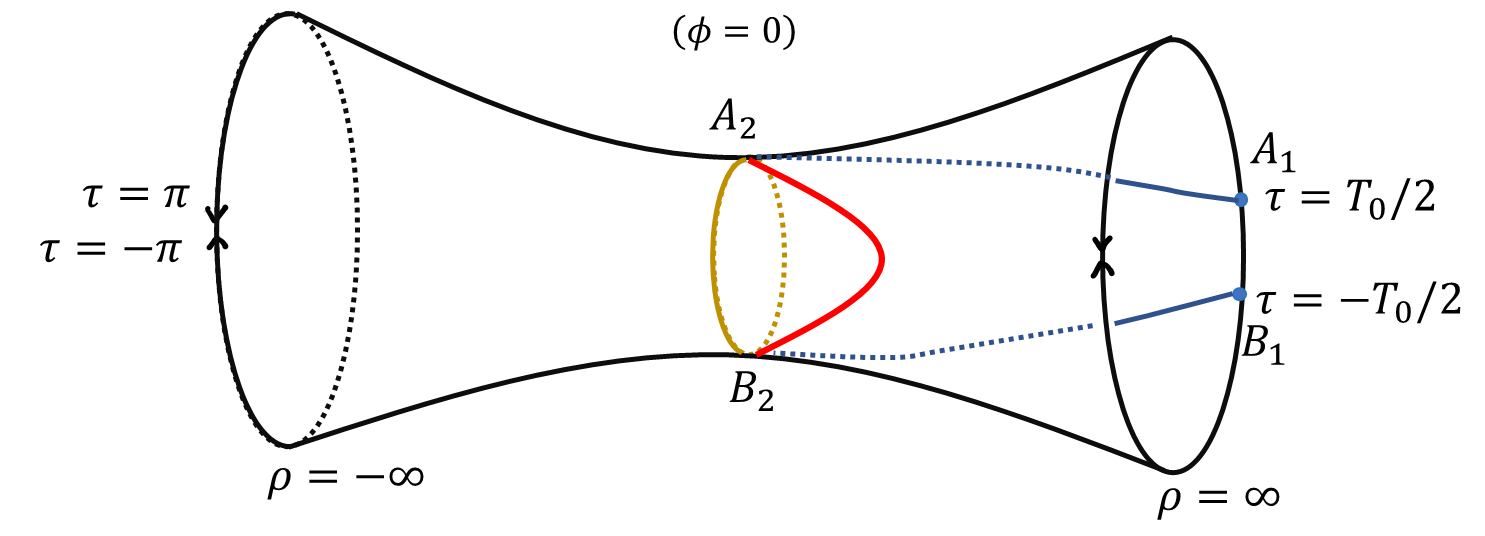}
   \caption{The configurations of CWES of time-like interval in global AdS$_3$  spacetime. The blue curves and red curve stand for space-like and time-like segments of CWES of global AdS$_3$ spacetime. The golden circle stands for the another time-like geodesic locates at the throat $\rho=0$. }\label{figads3d}
\end{figure}
Due to the topology now is $S^1\times R^2$, expect for the CWES obtain from simple connected configuration, there are other configurations of CWES. In the Fig.~\ref{figads3d}, the CWES contains an other closed circle (denoted by golden curve) which locates at the  throat $\rho=0$. The circumference of this circle is $2\pi$. We can add arbitrary copies of this circle into our configuration of CWES. The general CWESs will be given by
$$(\text{simply connected CWES})\cup S_1^1\cup S_2^1\cdots\cup S_n^1\,.$$
Here $\{S_1^1, S_2^1, \cdots, S_n^1\}$ stand for different closed timelike geodesics which lay on the hypersurface $\phi=0$ and are homotopic to the throat $\{\rho=0\}$. Their complex-valued area reads
\begin{equation}\label{totalAads3b}
  \mathscr{A}_n=2R_{\text{AdS}}\ln\frac{2\sin(T_0/2)}{\epsilon}+i(1+2n)R_{\text{AdS}}\pi,~~~n=0,1,2,\cdots\,.
\end{equation}
From our holographic proposal~\eqref{214}, we obtain holographic time-like entanglement entropy
\begin{equation}
	\label{findglobalads3}
\begin{split}
	S_{A}&=\frac{c_{\text{AdS}}}6\mathrm{Min}\left\{2\ln\frac{2\sin(T_0/2)}{\epsilon}+i(1+2n)\pi,~~~n=0,1,2,\cdots\right\}\\
&=\frac{c_{\text{AdS}}}3\ln\frac{2\sin(T_0/2)}{\epsilon}+\frac{ic_{\text{AdS}}\pi}6.
\end{split}
\end{equation}
We then recover the result obtained by Ref.~\cite{Doi:2022iyj}. For a general spacelike interval with endpoints $(0,0)$ and $(X,T_0)$ in global AdS$_3$, the entanglement entropy has following expression
\begin{equation}
	\label{ads3sa1}
	S_{A}=\frac{c_{\text{AdS}}}6\ln\left[\frac{4}{\epsilon^2}\left(\sinh^2\frac{X}{2}-\sin^2\frac{T_0}{2}\right)\right]\,.
\end{equation}
Now take $X=0$ and one will obtain
\begin{equation}
	\label{ads3sa2}
	S_{A}=\frac{c_{\text{AdS}}}6\left[2\ln\frac{2\sin(T_0/2)}{\epsilon}+i(1+2n)\pi\right],~~~n=0,1,2,\cdots\,.
\end{equation}
Interestingly, the result~\eqref{totalAads3b} also just corresponds to the multiple values~\eqref{ads3sa2} when we analytically continue logarithmic function into complex plane.

In order to exhibit how to use our proposal to obtain the correct result step by step, above computations look lengthy. However, one can clearly see that all above computations are only based on our holographic proposal~\eqref{214} itself without referring to the analytical continuation of space-like intervals. In following, we will discuss more examples to support our proposal.


\section{More examples}\label{moreexamples}
\subsection{AdS$_{3}$/CFT$_{2}$ with two intervals}
\begin{figure}
\centering
\subfigure[]{
\includegraphics[width=0.3\textwidth]{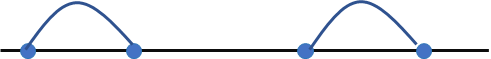}}
\subfigure[]{
\includegraphics[width=0.3\textwidth]{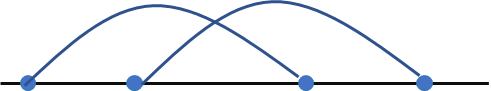}}
\subfigure[]{
\includegraphics[width=0.3\textwidth]{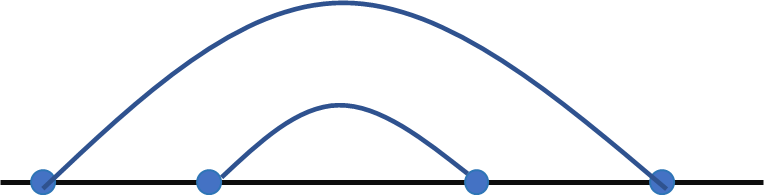}}
   \caption{Three different possible configurations of extremal surfaces for two spacelike intervals in planar CFT$_2$.}\label{twointerval1}
\end{figure}

In this section, let us consider a little complicated case: in the boundary of Poincar\'{e} patch, the timelike subregion contains two disjointed intervals $A=[-T_0-l/2,-l/2]\cup[l/2,T_0+l/2]$. We first consider its spacelike partner, i.e. two disconnected spacelike intervals $[-L-l/2,-l/2]\cup[l/2,L+l/2]$. From holography we find we need to consider three different possible configurations shown in the Fig.~\ref{twointerval1}(a), (b) and (c), of which the holographic entanglement entropies are easy to find by the geodesics of AdS$_3$ spacetime,
\begin{equation}\label{timelkies1}
  \mathscr{A}(a)=2 R_{\text{AdS}}\ln \left(\frac{L^2}{\epsilon^2}\right),~~\mathscr{A}(b)=2R_{\text{AdS}} \ln \left(\frac{(L+l)^2}{\epsilon^2}\right), ~~\mathscr{A}(c)=2R_{\text{AdS}}\ln \left(\frac{(2L+l)l}{\epsilon^2}\right)
\end{equation}
It is clear that $\mathscr{A}(a)\leq \mathscr{A}(b)$ and $\mathscr{A}(c)\leq \mathscr{A}(b)$, so we only need to compare the configurations (a) and (c). Define
$$x=\frac{(2L+l)l}{L^2}$$
and we have following result for spacelike two disjointed intervals
\begin{equation}\label{timelkies2}
  \text{Min}\mathscr{A}=\left\{
  \begin{split}
  &4R_{\text{AdS}} \ln \left(\frac{L}{\epsilon}\right),\text{if}~x>1\\
  &2R_{\text{AdS}}\ln \left(\frac{(2L+l)l}{\epsilon^2}\right),~~~0<x<1
  \end{split}
    \right.
\end{equation}
The timelike entanglement entropy then is easy to obtained by analytical condition $L\rightarrow iT_0$ and $l\rightarrow il$, which gives us
\begin{equation}\label{timelkies2}
  \text{Min}\mathscr{A}=\left\{
  \begin{split}
  &4 R_{\text{AdS}}\ln \left(\frac{T_0}{\epsilon}\right)+2i\pi R_{\text{AdS}},\text{if}~x>1\\
  &2R_{\text{AdS}}\ln \left(\frac{(2T_0+l)l}{\epsilon^2}\right)+2i\pi R_{\text{AdS}},~~~0<x<1
  \end{split}
    \right.
\end{equation}
Here $x=(2T_0+l)l/T_0^2$.

\begin{figure}
\centering
\subfigure[]{
\includegraphics[width=0.2\textwidth]{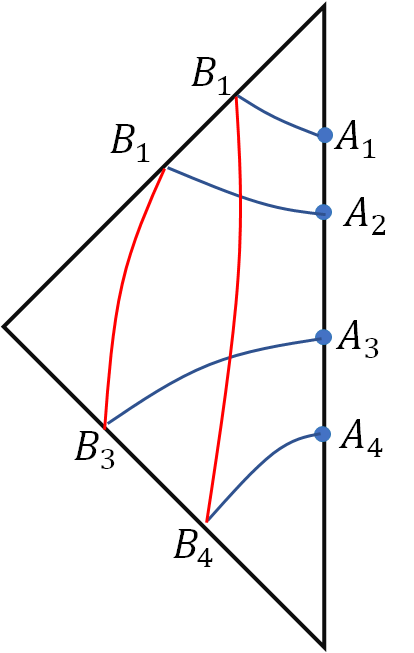}}
\subfigure[]{
\includegraphics[width=0.2\textwidth]{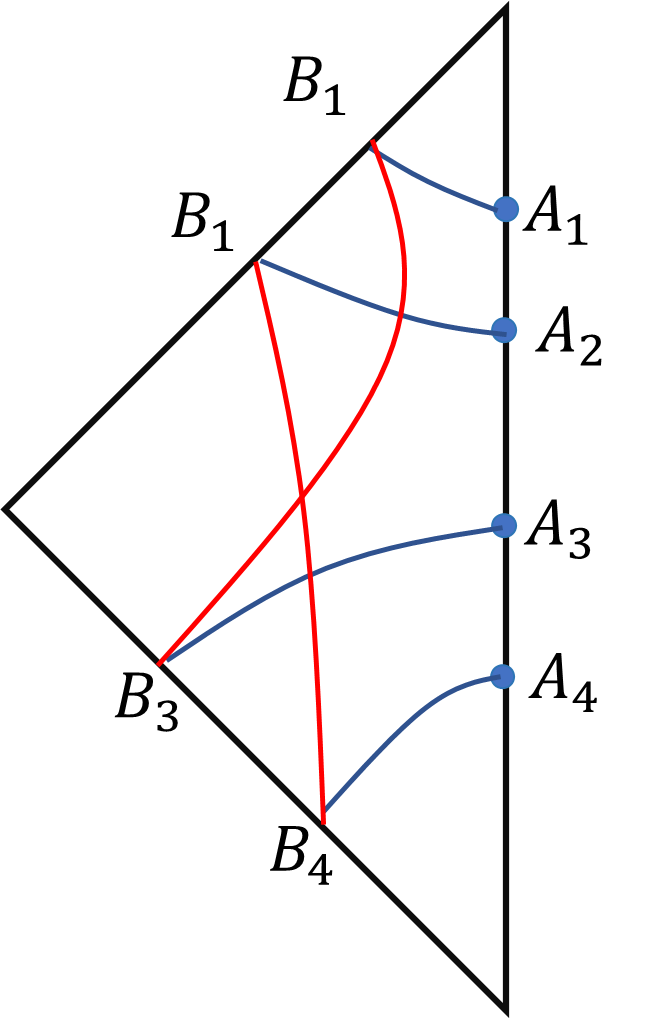}}
\subfigure[]{
\includegraphics[width=0.2\textwidth]{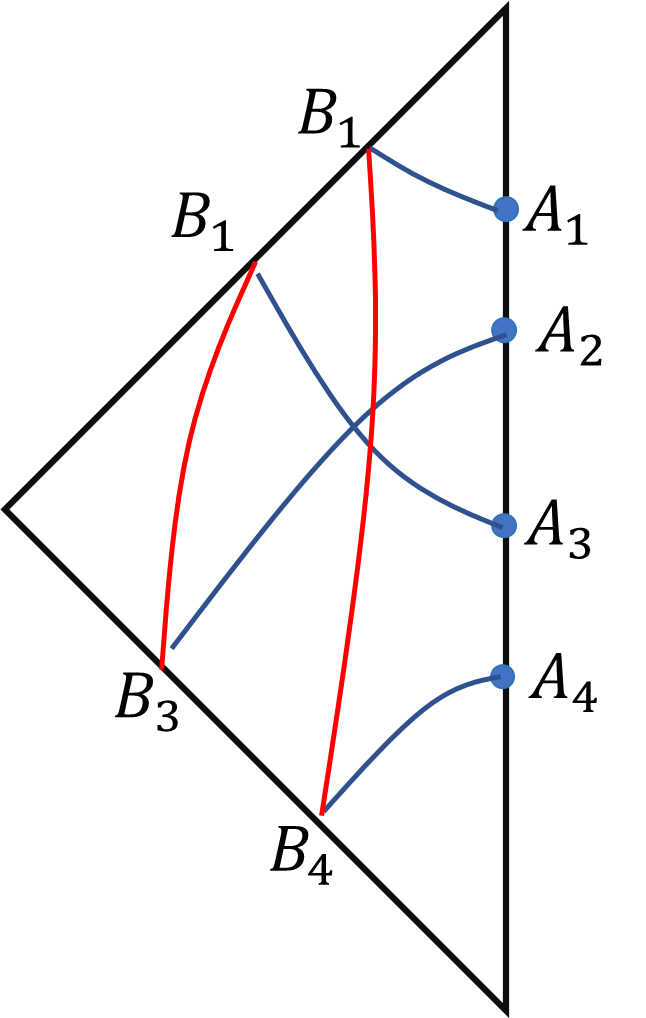}}
\subfigure[]{
\includegraphics[width=0.2\textwidth]{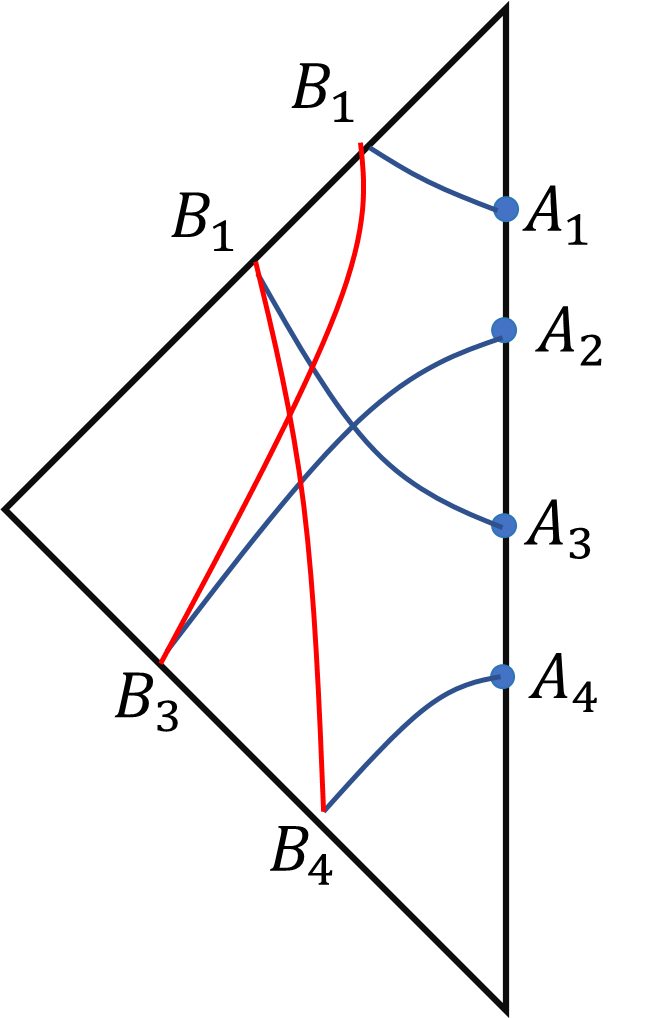}}
   \caption{Four of 12 different possible configurations of geodesics for two disjointed timelike intervals. }\label{twoint1}
\end{figure}

Now let us consider the two timelike disjointed intervals by using our holographic proposal. There are 4 endpoints at the line $x=0$ of AdS boundary. Two of them are connected to future null infinity by spacelike geodesics. We then at most have total $2\times C_4^2=12$ different possible configurations. Four of them are shown in the Fig.~\ref{twoint1}.
Using a similar argument of the discussion on the Poincar\'{e} patch of AdS$_3$ spacetime, one can find that configuration (c) cannot be a CWES. Using the result of Eq.~\eqref{totalAads3} we can find the area of corresponding CWES in the other three different configurations are
\begin{equation}\label{twointers1}
  \mathscr{A}(a)=2R_{\text{AdS}}\ln\frac{(2T_0+l)l}{\epsilon^2}+2iR_{\text{AdS}}\pi,~~\mathscr{A}(b)=2R_{\text{AdS}}\ln\frac{(T_0+l)^2}{\epsilon^2}+2iR_{\text{AdS}}\pi
\end{equation}
and
\begin{equation}\label{twointers1}
  \mathscr{A}(d)=2R_{\text{AdS}}\ln\frac{T_0^2}{\epsilon^2}+2iR_{\text{AdS}}\pi\,.
\end{equation}
The other 8 different configurations will also give above 3 different results and so our holographic proposal recovers Eq.~\eqref{timelkies2}.

\subsection{BTZ black holes}
We now consider the time-like entanglement entropy in finite temperature system. The the gravity dual is given by the
BTZ black hole
\begin{equation}\label{btzg1}
  \td s^2=R_{\text{AdS}}^2\left[-(r^2-r_h^2)\td t^2+\frac{\td r^2}{r^2-r_h^2}+r^2\td x^2\right]
\end{equation}
where $r_h=2\pi/\beta$ and $\beta$ is the inverse temperature of dual CFT. When A is a time-like interval with length $T_0$, the time-like entanglement entropy has been found~\cite{Doi:2022iyj}
\begin{equation}
	\label{ftebtz1}
	S_{A}=\frac{c_{\text{AdS}}}3\ln\left[\frac{\beta}{\pi\epsilon}\sinh\frac{\pi T_0}{\beta}\right]+\frac{ic_{\text{AdS}}\pi}6\,.
\end{equation}
Before we check our holographic proposal, let us first simply mention how to obtain this result from analytical continuation. We know that for a space-like interval of finite length $L$ in thermal 2-dimensional CFT, then entanglement entropy reads
\begin{equation}
	\label{ftebtz2}
	S_{A}=\frac{c_{\text{AdS}}}3\ln\left[\frac{\beta}{\pi\epsilon}\sinh\frac{\pi L}{\beta}\right]\,.
\end{equation}
Leaning from Eq.~\eqref{201}, naively one may think the time-like entanglement entropy is obtained by a simple replacement $L\rightarrow iT_0$. If we do so, we will then obtain
\begin{equation}
	\label{ftebtz3}
	S_{A}=\frac{c_{\text{AdS}}}3\ln\left[\frac{\beta}{\pi\epsilon}\sin\frac{\pi T_0}{\beta}\right]+\frac{ic_{\text{AdS}}\pi}6\,.
\end{equation}
We see this is different from Eq.~\eqref{ftebtz1}. Eq.~\eqref{ftebtz3} is obvious wrong since the length $T_0$ of time-like interval in a BTZ black hole can run from 0 to $\infty$. The reason is that formula~\eqref{ftebtz2} is the result of space-like interval in static slice, i.e. the straight line from $(0,0)$ to $(0,L)$ in the $(t,x)$ plane, rather than the entanglement entropy of arbitrary space-like interval. To obtain its analytical continuation of time-like interval, we should obtain an analytical expression for entanglement entropy for the straight line from $(0,0)$ to $(t,L)$ in the $(t,x)$ plane. Even in CFT$_2$-BTZ black hole case, such computation is still very complex. Fortunately,  we obtain such analytical result,
\begin{equation}
	\label{ftebtz4}
	S_{A}=\frac{c_{\text{AdS}}}6\ln\left[\frac{\beta^2}{\pi^2\epsilon^2}\left(\sinh^2\frac{\pi L}{\beta}-\sinh^2\frac{\pi t}{\beta}\right)\right]\,.
\end{equation}
We obtain this formula by follow facts. Firstly we know that two point function of Euclidean BTZ black hole reads
\begin{equation}
	\label{ftebtz4a}
	G_E=\left(\sinh^2\frac{\pi L}{\beta}+\sin^2\frac{\pi t_E}{\beta}\right)^{-\Delta}\,.
\end{equation}
Here $\Delta$ is the conformal dimensional of operator and $t_E$ is the Euclidean time. From this result one can use replica trick and twist operators to obtain the entanglement entropy of straight line from $(0,0)$ to $(t,L)$ in the Euclidean $(t_E,x)$ plane
\begin{equation}
	\label{ftebtz4b}
	S_{A}=\frac{c_{\text{AdS}}}6\ln\left[\frac{\beta^2}{\pi^2\epsilon^2}\left(\sinh^2\frac{\pi L}{\beta}+\sin^2\frac{\pi t_E}{\beta}\right)\right]\,.
\end{equation}
Now Wick rotate into Lorentz signature and we obtain Eq.~\eqref{ftebtz4}. Note that formula~\eqref{ftebtz4} is obtained by assume $L>t>0$. We now analytically continue it into time-like interval and set $L=0, t=T_0$, which gives us the correct result~\eqref{ftebtz1}. This example also tell us why it may be difficult to use analytical continuation to compute the time-like entanglement entropy: in order to obtain the correct analytical continuation, we in general need to find the entanglement entropy for arbitrary space-like interval, such as formula~\eqref{ftebtz4}, but this is very difficult in general case. For example, for a thermal CFT$_d$ with $d\geq3$, which is dual to a $(d+1)$-dimensional planar Schwarzschild AdS black brane, there is no analytical formula about the entanglement entropy of general space-like subregions.

\begin{figure}
\centering
\subfigure[]{
\includegraphics[width=0.35\textwidth]{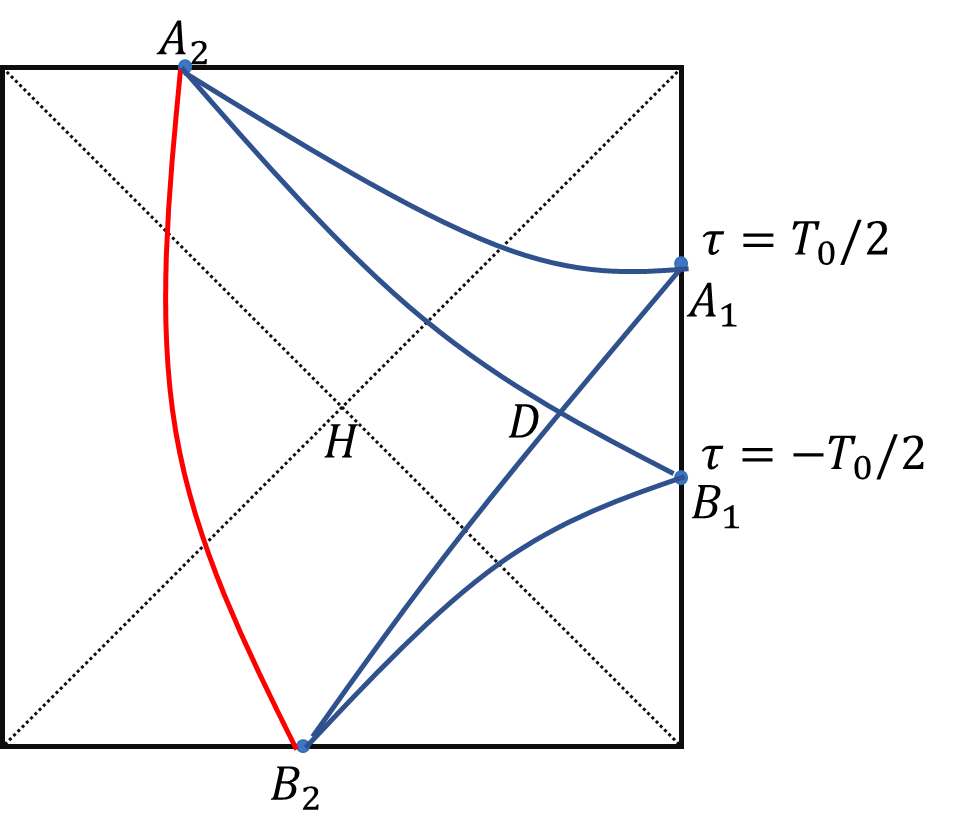}}
\subfigure[]{
\includegraphics[width=0.35\textwidth]{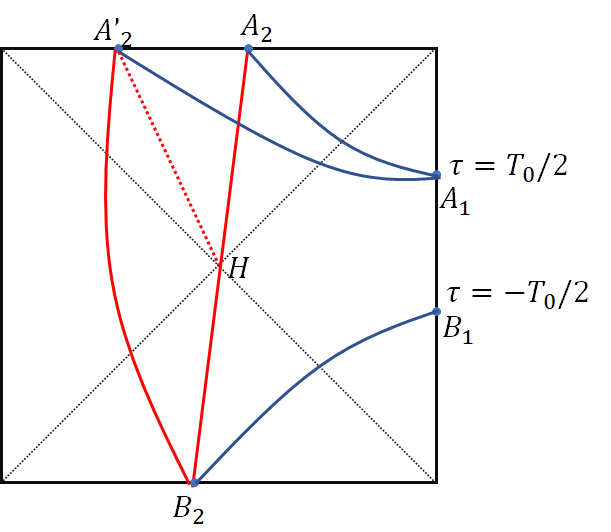}}
   \caption{The configurations of CWES of time-like interval in BTZ black hole. The blue curves stand for space-like geodesics and red curves stand for time-like geodesics. Panel a: There are two types of relevant configurations of CWESs, which are  $A_1A_2B_2B_1$ and $A_1B_2A_2B_1$. Panel b: the time-like geodesic $A_2B_2$ passes through the bifurcated ``point'' of event horizon. }\label{figbtz1}
\end{figure}

Now let us explain how to use our holographhic proposal to recover Eq.~\eqref{ftebtz1}. We denote the endpoints of time-like interval $A$ to be $A_1$ and $B_2$ with the time coordinate $t=\pm T_0/2$. In Fig.~\ref{figbtz1}(a) we exhibit schematically the two relevant configurations when we try to find the minimal CWES. The one configuration is the curve $A_1A_2B_2B_1$ and the other one is the curve $A_1B_2A_2B_1$. For space-like geodesics, we have ``triangle inequality'', which tells us
\begin{equation}\label{spaceineq}
  \mathscr{A}(A_1A_2)<\mathscr{A}(A_1D)+\mathscr{A}(DA_2),~~\mathscr{A}(B_1B_2)<\mathscr{A}(B_1D)+\mathscr{A}(DB_2)
\end{equation}
Thus, we see
\begin{equation}\label{spaceineq2}
  \mathscr{A}(A_1A_2B_2B_1)\prec\mathscr{A}(A_1B_2A_2B_1)\,.
\end{equation}
by our rule~\eqref{213}. We then conclude that, to find the minimal CWES we only need to consider the configuration $A_1A_2B_2B_1$ of Fig.~\ref{figbtz1}(a).

There are two different cases shown in Fig.~\ref{figbtz1}(b). The red line $A_2B_2$ is the time-like geodesic passing through the bifurcated ``point'' $H$ of event horizon. the red curve $A'_2B_2$ is a general time-like geodesic connecting the points of past and future singularity. Denote the straight line $A_2'H$ to be timlike geodesic connecting $A_2'$ and $H$. Since timlike geodesics obey the ``anti-triangle inequality'', we then find
\begin{equation}\label{spaceineq3a}
  \mathscr{A}(A_2'B_2)\succ\mathscr{A}(A_2'H)+\mathscr{A}(HB_2)
\end{equation}
On the other hand, the time translation symmetry shows that $\mathscr{A}(A_2'H)=\mathscr{A}(A_2H)$. We then see
\begin{equation}\label{spaceineq2}
  \mathscr{A}(A_1A_2B_2B_1)\prec\mathscr{A}(A_1A'_2B_2B_1)\,.
\end{equation}
This means that, to find the minimal CWES, we only need to study the configuration shown by the curve $A_1A_2B_2B_1$ of Fig.~\ref{figbtz1}(b).

Since the time-like geodesic passes through the bifurcated ''point'', its length is constant and we obtain
\begin{equation}\label{spaceineq2}
  \text{Im}\mathscr{A}(A_1A_2B_2B_1)=R_{\text{AdS}}\pi\,.
\end{equation}
To find the real part of $\mathscr{A}(A_1A_2B_2B_1)$, we denote the position of $A_2$ and $B_2$ to be
$$A_2=(r=0, \tau=\tau_0),~~B_2=(r=0, \tau=-\tau_0)\,.$$
Since the joint points are at boundary of spacetime, the curve $A_1A_2B_2B_1$ is a the minimal CWES only if following equation
\begin{equation}\label{condibtz1}
  \frac{\partial}{\partial \tau_0}\text{Re}\mathscr{A}(A_1A_2B_2B_1)=0\,.
\end{equation}
is satisfied. In order to describe the geodesic $A_1A_2$ of Fig.~\ref{figbtz1}(b), we transform into Eddington coordinate
\begin{equation}\label{btzg1}
  \td s^2=R_{\text{AdS}}^2[-(r^2-r_h^2)\td u^2+2\td u\td r+r^2\td x^2]\,.
\end{equation}
Where $u$ and $t$ are transformed by
$$u=t+\frac1{2r_h}\ln\left|\frac{r-r_h}{r+r_h}\right|\,.$$
The endpoints of $A_1A_2$ then is $(r=0, u=\tau_0)$ and $(r=\infty, u=T_0/2)$. To be space-like geodesic we should restrict
$$\tau_0<T_0/2\,.$$
The geodesic $A_1A_2$ is given by following equation
\begin{equation}\label{geodeqsbtz}
  r''-3rr'+r(r^2-r_h^2)=0\,.
\end{equation}
Here the prime stands for the derivative with respective to $u$. The solution to describe geodesic $A_2A_1$ reads
\begin{equation}\label{solutbtz1}
  r(u) = \frac{r_h\sinh(r_h(u-\tau_0))}{\cosh(r_h(T_0/2-\tau_0)-\cosh(r_h(u-\tau_0))},~~u\in[\tau_0, T_0/2)\,.
\end{equation}
Its geodesic length reads
\begin{equation}\label{legtha1a2}
  \mathscr{A}(A_1A_2)=R_{\text{AdS}}\ln\left[\frac{\beta}{\pi\epsilon}\sinh\frac{\pi (T_0-2\tau_0)}{\beta}\right]\,.
\end{equation}
Similarity we can find
\begin{equation}\label{legtha1a3}
  \mathscr{A}(B_1B_2)=R_{\text{AdS}}\ln\left[\frac{\beta}{\pi\epsilon}\sinh\frac{\pi (T_0+2\tau_0)}{\beta}\right]\,.
\end{equation}
Thus we find that the real part of $\mathscr{A}(A_1A_2B_2B_1)$
\begin{equation}\label{legtha1a4}
  \text{Re}\mathscr{A}(A_1A_2B_2B_1)=2R_{\text{AdS}}\ln\left[\frac{\beta}{\pi\epsilon}\sqrt{\sinh\frac{\pi (T_0+2\tau_0)}{\beta}\sinh\frac{\pi (T_0-2\tau_0)}{\beta}}\right]\,.
\end{equation}
Eq.~\eqref{condibtz1} then gives us $\tau_0=0$ and so we find area of the CWES reads
\begin{equation}\label{legtha1a4b}
 \mathscr{A}(A_1A_2B_2B_1)=2R_{\text{AdS}}\ln\left[\frac{\beta}{\pi\epsilon}\sinh\frac{\pi T_0}{\beta}\right]+iR_{\text{AdS}}\pi
\end{equation}
Then we recover the result~\eqref{ftebtz1}.

\begin{figure}
\centering
\includegraphics[width=0.35\textwidth]{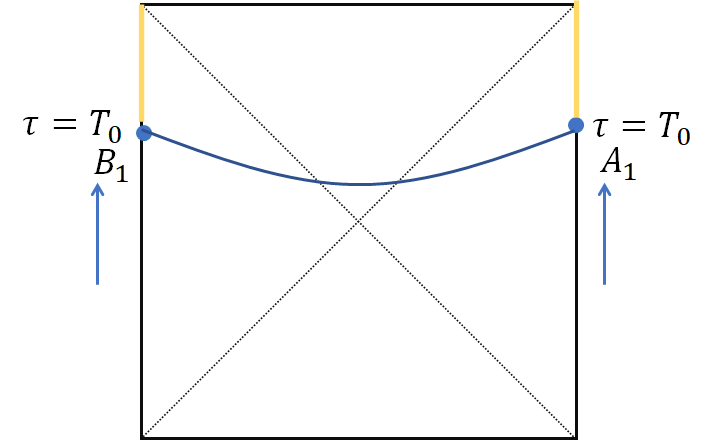}
   \caption{The two space-like half lines are denoted by blue dots and the  two time-like half lines are denoted by golden solid lines.  The arrows indicate the time direction of two boundaries.}\label{figbtz3}
\end{figure}
The maximally analytically continued Penrose diagram of BTZ black hole contains two-copy of the boundaries. These two boundaries correspond to two copies of the field theory. Such eternal black hole in holography is conjectured dual to the thermofield double states~\cite{Maldacena:2001kr}. In general, these two copies are in an entangled with each others. When we consider the  thermofield double states, the time direction of time boundaries are usually same, see Fig.~\ref{figbtz3}. We now consider the time-like entanglement entropy of these two boundary CFTs. The time-like interval $A$ then contains two half time lines $\tau\in(T_0,\infty)$ at two boundaries, see the golden lines of Fig.~\ref{figbtz3}. Even though we now consider the time-like entanglement entropy here, we note that there is a space-like geodesic connecting two endpoints of time-like interval $A$. Thus, based on our rule of minimal CWES, the time-like entanglement entropy is given by this space-like geodesic and we obtain a real-valued time-like entanglement entropy
\begin{equation}\label{btztfd0}
  S_A=\frac{c_{\text{AdS}}}3\ln\left[\frac{\beta}{2\pi\epsilon}\cosh\frac{2\pi T_0}{\beta}\right]\,.
\end{equation}
This is a little surprising and seems to be an incorrect result due to lack of imaginary part. However, this is indeed what we should obtain. To understand this point, consider two half space-like lines belong to the two boundary CFTs. When these two half lines are regular time slices with the endpoints $A_1$ and $B_1$. The entanglement entropy between them is given by~\cite{Hartman:2013qma}
\begin{equation}\label{btztfd1}
  S=\frac{c_{\text{AdS}}}3\ln\left[\frac{\beta}{2\pi\epsilon}\cosh\frac{2\pi T_0}{\beta}\right]\,.
\end{equation}
This can obtained by computing the space-like geodesics connecting $A_1$ and $B_1$. Though this equation is obtain by assuming that the half lines have constant time $t$, this result is still true for arbitrary space-like half lines of which the endpoints at two boundaries are also $A_1$ abd $B_1$. This shows that if we ``rotate'' the space-like half lines into time-like but fix the endpoints, the entropy should be keep invariant. Thus we still obtain Eq.~\eqref{btztfd0} for the time-like interval $A$. This shows our holographic proposal still gives the correct result. This example also shows that the time-like entanglement entropy is not always complex value.

\section{Summary and discussion}\label{dis}
In this work we consider time-like entanglement entropy proposed by Ref.~\cite{Doi:2022iyj}. This generalization in general will be complex-valued and can be interpreted as pseudo entropy. For some simple cases that we can obtain analytical expression of entanglement entropy for spacelike subregions, we can analytical continue spacelike regions into timelike regions and obtain the pseudo entropy. In order to obtain holographic interpretation on the complex-valued pseudo entropy of timelike subregions, Ref.~\cite{Doi:2022iyj} generalize the original smooth spacelike Ryu-Takayanagi extremal surface to mix space-like and time-like extremal surfaces. The complex-value pseudo entropies for various timelike intervals were first computed by analytical continuation and their holographic interpretations were given according to these results. However, if the holographic duality really works for the understanding of timelike entanglement entropy, we should have holographic proposal on how to find such complex-valued entropy for timelike subregions but do not need to refer to analytical continuation.   In this paper, we try to seek such a holographic proposal and unify the holographic (pseudo) entanglement entropy for both spacelike and timelike subregions.

Firstly, we use concrete example to show that, the naively combinations of spacelike and timelike extremal surfaces  will not be unique in general case. Even for the simplest case, i.e. the single timelike interval in planar CFT$_2$, we have infinitely many different choices of using spacelike and timelike geodesic in the bulk to connect the endpoints of the timelike interval. These different choices can give us infinitely many different complex-valued areas. Without referring to analytical continuation, there is not a priori method to help us pick up the correct one before this paper. To overcome this problem, this paper generalizes the conception of extremal surface of spacelike subregions and proposes the conception of CWES.  Our proposal naturally includes the holographic entanglement entropy of spacelike subregion as its special case. To justify the correctness of our proposal for timelike entanglement entropy, we give some concrete examples on how to use our proposal to find the unique holographic timlike entanglement entropy and show the results are as same as ones obtained from analytical continuations.

There are many questions which are in need of further study. Firstly, we only used our holographic proposal to study a few of simple situations that the result can be double-checked from analytical continuations. From the viewpoint of practice, the real application of our holographic proposal should be the case where analytical continuation is hard or even impossible to perform. For example, the Schwarzschild-AdS black hole and Reissner-Nordstr\"{o}m-AdS black hole. Does our proposal also give physically correct results? Secondly, our proposal is obtained just by guesswork. The holographic entanglement entropy of spacelike regions, i.e. the RT formula, can be proved holographically by using holographic replica trick~\cite{Lewkowycz:2013nqa,Dong:2016fnf} or other method~\cite{Casini:2011kv}. If our proposal of holographic timelike entanglement entropy is really correct, could we find any holographic method to prove it? Thirdly, it was proposed that the two-point function of heavy operators can be computed holographically by ``geodesic approximation''~\cite{Balasubramanian:1999zv}. In Lorentz signature, due to the lack of smooth geodesics to connect the two timelike points, such method cannot be applied into the computations of two timelike points. However, inspired by the work of Ref.~\cite{Doi:2022iyj}, one may also consider the geodesics of mixing spacelike and timelike segments to generalize the ``geodesic approximation''.

What's more, from our computation of BTZ black hole, one can realize that the holographic timelike entanglement entropy can probe the interior information behind event horizon. In the previous studies such as entanglement entropy or complexity~\cite{Brown:2015bva,Brown:2015lvg,Alishahiha:2015rta,Carmi:2016wjl}, we have to use both two boundaries of an eternal black hole to probe the interior behind horizon. Such a method can only be applied into eternal black holes. From Fig.~\ref{figbtz1} we see that the extremal surfaces of a timelike subregion can entry into interior of horizon. It is not difficultly to realize such phenomenon will also happen in general eternal black holes. Thus, the study on timelike entanglement entropy may open a window on how to use such a quantum information tool to probe and reconstruct interior of black hole from its single boundary. This has practical significance since the black hole formed by collapse will have only single boundary in its Penrose diagram.

\acknowledgments
This work is supported by the National Natural Science Foundation of China under Grant No. 12005155.

\appendix

\bibliographystyle{JHEP}

\bibliography{ref-WeakExtremalSurface-1}

\end{document}